\documentclass[longauth]{aa}  
\usepackage{graphicx}
\usepackage{txfonts,textcomp}
\usepackage{lipsum}
\usepackage[dvipsnames]{xcolor}
\usepackage{stfloats}
\usepackage{multirow}
\usepackage[version=4]{mhchem}
\usepackage[utf8]{inputenc}
\usepackage{booktabs}
\usepackage{amssymb}
\usepackage{caption} 
\usepackage{xhfill}
\usepackage{array}
\usepackage{hyperref}
\usepackage[euler]{textgreek}
\usepackage{natbib}
\usepackage{subfig}
\bibpunct{(}{)}{;}{a}{}{,}
\usepackage{soul}

\newcolumntype{L}[1]{>{\raggedright\let\newline\\\arraybackslash\hspace{0pt}}m{#1}}
\newcolumntype{C}[1]{>{\centering\let\newline\\\arraybackslash\hspace{0pt}}m{#1}}
\newcolumntype{R}[1]{>{\raggedleft\let\newline\\\arraybackslash\hspace{0pt}}m{#1}}

\newcommand*{\dprime}{^{\prime\prime}\mkern-1.2mu}
\newcommand{\micron}{\textmu m\xspace}

\definecolor{darkorange}{rgb}{1.0, 0.55, 0.0}

\newcommand{\orcidlink}[1]{\protect\href{https://orcid.org/#1}{\protect\includegraphics[width=8pt]{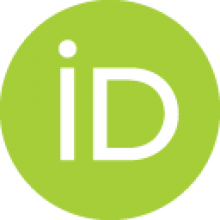}}} 

\begin{document} 
    \title{A JWST/MIRI analysis of the ice distribution and PAH emission in the protoplanetary disk HH~48~NE}

   \author{ J.~A.~Sturm              \inst{1}\thanks{sturm@strw.leidenuniv.nl}\orcidlink{0000-0002-0377-1316}       \and
            M.~K.~McClure            \inst{1}\orcidlink{0000-0003-1878-327X}                                        \and
            D.~Harsono               \inst{2}\orcidlink{0000-0001-6307-4195}                                        \and
            J.~B.~Bergner            \inst{3}\orcidlink{0000-0002-8716-0482}                                        \and
            E.~Dartois               \inst{4}\orcidlink{0000-0003-1197-7143}                                        \and
            A.~C.~A.~Boogert         \inst{5}\orcidlink{0000-0001-9344-0096}                                        \and  
            M.~A.~Cordiner           \inst{6,7}\orcidlink{0000-0001-8233-2436}                                      \and  
            M.~N.~Drozdovskaya       \inst{8}\orcidlink{0000-0001-7479-4948}                                        \and
            S.~Ioppolo               \inst{9}\orcidlink{0000-0002-2271-1781}                                        \and
            C.~J.~Law                \inst{10}\thanks{NASA Hubble Fellowship Program Sagan Fellow}\orcidlink{0000-0003-1413-1776}                                       \and
            D.~C.~Lis                \inst{11}\orcidlink{0000-0002-0500-4700}                                       \and
            B.~A.~McGuire            \inst{12,13}\orcidlink{0000-0003-1254-4817}                                    \and
            G.~J.~Melnick            \inst{14}\orcidlink{0000-0002-6025-0680}                                       \and
            J.~A.~Noble              \inst{15}\orcidlink{0000-0003-4985-8254}                                       \and
            K.~I.~\"Oberg            \inst{14}\orcidlink{0000-0001-8798-1347}                                       \and   
            M.~E.~Palumbo            \inst{16}\orcidlink{0000-0002-9122-491X}                                       \and
            Y.~J.~Pendleton          \inst{17}\orcidlink{0000-0001-8102-2903}                                       \and
            G.~Perotti               \inst{18}\orcidlink{0000-0002-8545-6175}                                       \and
            W.~R.~M.~Rocha           \inst{1,19}\orcidlink{0000-0001-6144-4113}                                     \and
            R.~G.~Urso               \inst{16}\orcidlink{0000-0001-6926-1434}                                       \and
            E.~F.~van~Dishoeck       \inst{1,20}\orcidlink{0000-0001-7591-1907}                                       
          }

    \institute{
    Leiden Observatory, Leiden University, P.O. Box 9513, NL-2300 RA Leiden, The Netherlands                                                        \and
    Institute of Astronomy, Department of Physics, National Tsing Hua University, Hsinchu, Taiwan                                                   \and
    Department of Chemistry, University of California, Berkeley, California 94720-1460, United States                                               \and
    Institut des Sciences Mol\'eculaires d’Orsay, CNRS, Univ. Paris-Saclay, 91405 Orsay, France                                                     \and
    Institute for Astronomy, University of Hawai'i at Manoa, 2680 Woodlawn Drive, Honolulu, HI 96822, USA                                           \and
    Astrochemistry Laboratory, NASA Goddard Space Flight Center, 8800 Greenbelt Road, Greenbelt, MD 20771, USA                                      \and
    Department of Physics, Catholic University of America, Washington, DC 20064, USA                                                                \and
    Physikalisch-Meteorologisches Observatorium Davos und Weltstrahlungszentrum (PMOD/WRC), Dorfstrasse 33, CH-7260, Davos Dorf, Switzerland        \and
    Centre for Interstellar Catalysis, Department of Physics and Astronomy, Aarhus University, DK 8000 Aarhus, Denmark                              \and
    Department of Astronomy, University of Virginia, Charlottesville, VA 22904, USA                                                                 \and
    Jet Propulsion Laboratory, California Institute of Technology, 4800 Oak Grove Drive, Pasadena, CA, 91109, USA                                   \and
    Department of Chemistry, Massachusetts Institute of Technology, Cambridge, MA 02139, USA                                                        \and
    National Radio Astronomy Observatory, Charlottesville, VA 22903, USA                                                                            \and
    Center for Astrophysics \textbar\ Harvard \& Smithsonian, 60 Garden St., Cambridge, MA 02138, USA                                               \and
    Physique des Interactions Ioniques et Mol\'{e}culaires, CNRS, Aix Marseille Univ., 13397 Marseille, France                                      \and
    INAF – Osservatorio Astrofisico di Catania, via Santa Sofia 78, 95123 Catania, Italy                                                            \and
    Department of Physics, University of Central Florida, Orlando, FL 32816, USA                                                                    \and
    Max Planck Institute for Astronomy, K{\"o}nigstuhl 17, D-69117 Heidelberg, Germany                                                              \and
    Laboratory for Astrophysics, Leiden Observatory, Leiden University, PO Box 9513, 2300 RA Leiden, The Netherlands                                \and
    Max-Planck-Institut f{\"u}r extraterrestrische Physik, Giessenbachstra{\ss}e 1, 85748 Garching bei M\"unchen, Germany 
    }

    \date{Received XXX; accepted YYY}
    \abstract
    {Ice-coated dust grains provide the main reservoir of volatiles that play an important role in planet formation processes and may become incorporated into planetary atmospheres. 
    However, due to observational challenges, the ice abundance distribution in protoplanetary disks is not well constrained. 
    With the advent of the \textit{James Webb} Space Telescope (JWST) we are in a unique position to observe these ices in the near- to mid-infrared and constrain their properties in Class~II protoplanetary disks.
    }
    {We present JWST Mid-InfraRed Imager (MIRI) observations of the edge-on disk HH~48~NE carried out as part of the Director’s Discretionary Early Release Science program Ice Age, completing the ice inventory of HH~48~NE by combining the MIRI data (5~--~28~\micron) with those of NIRSpec (2.7~--~5~\micron). 
    }
    {We used radiative transfer models tailored to the system, including silicates, ices, and polycyclic aromatic hydrocarbons (PAHs) to reproduce the observed spectrum of HH~48~NE with a parameterized model. 
    The model was then used to identify ice species and constrain spatial information about the ices in the disk.
    }
    {The mid-infrared spectrum of HH~48~NE is relatively flat with weak ice absorption features. 
    We detect \ce{CO2}, \ce{NH3}, \ce{H2O} and tentatively \ce{CH4} and \ce{NH4+}.
    Radiative transfer models suggest that ice absorption features are produced predominantly in the 50~--~100~au region of the disk. The \ce{CO2} feature at 15~\micron probes a region closer to the midplane ($z/r$ = 0.1~--~0.15) than the corresponding feature at 4.3~\micron ($z/r$ = 0.2~--~0.6), but all observations trace regions significantly above the midplane reservoirs where we expect the bulk of the ice mass to be located.
    Ices must reach a high scale height ($z/r~\sim 0.6$; corresponding to modeled dust extinction $A_{\rm v}~\sim 0.1$), in order to be consistent with the observed vertical distribution of the peak ice optical depths.
    The weakness of the \ce{CO2} feature at 15~\micron relative to the 4.3~\micron feature and the red emission wing of the 4.3~\micron \ce{CO2} feature are both consistent with ices being located at high elevation in the disk.
    The retrieved \ce{NH3} abundance and the upper limit on the \ce{CH3OH} abundance relative to \ce{H2O} are significantly lower than those in the interstellar medium (ISM), but consistent with cometary observations.
    The contrast of the PAH emission features with the continuum is stronger than for similar face-on protoplanetary disks, which is likely a result of the edge-on system geometry. 
    Modeling based on the relative strength of the emission features suggests that the PAH emission originates in the disk surface layer rather than the ice absorbing layer.}
    {Full wavelength coverage is required to properly study the abundance distribution of ices in disks.
    To explain the presence of ices at high disk altitudes, we propose two possible scenarios: a disk wind that entrains sufficient amounts of dust, thus blocking part of the stellar UV radiation, or vertical mixing that cycles enough ices into the upper disk layers to balance ice photodesorption from the grains.
    }
    
    \maketitle

\section{Introduction}\label{sec:introduction}
The composition of planetesimals, comets, and eventually planets is determined in large part by the composition of their building blocks: ice-coated dust grains. 
Ices are the dominant carriers of volatiles in planet-forming regions \citep{Pontoppidan2014, Walsh2015}, and set for a large part the spatial distribution of volatiles in the disk \citep{Mcclure2019,Banzatti2020,Sturm2022CI,Banzatti2023}.
Ices not only play a crucial role in the disk chemistry, but also in planet formation processes \citep[e.g.,][]{Oberg2016, Drazkowska2016}.
Unlike the gas present in the protoplanetary disk, ices are directly incorporated into the cores of planets, comets, and icy moons \citep{Oberg2023}.
Mapping ices in disks at different stages of their evolution is therefore important to understand the initial building blocks available for planet formation.

With the advent of the \textit{James Webb} Space Telescope (JWST) we are in the unique position to target the ice absorption bands in the mid-infrared (2~--~15~\micron) with sufficient angular resolution to search for spatial variations in the distribution of protoplanetary disk ices in nearby star-forming regions.
In the earlier stages of the star formation sequence, when the natal cloud or envelope has high line-of-sight $A_\mathrm{V}$ (Molecular clouds, Class~0 protostars, Class~I protostars), ices have been readily observed in detail with ISO, \textit{Spitzer} \citep{Boogert2015}, Akari \citep{Aikawa2012,Noble2013}, ground based observatories, and JWST \citep{Yang2022,McClure2023,Rocha2024,Brunken2024}.
However, in late-stage protoplanetary disks after the envelope has dissipated (Class~II systems), the contrast of these ice species with respect to the continuum from the warm inner disk is only detectable if the star and bright inner disk is blocked by the disk.
This is the case for edge-on disks, which makes them ideal laboratories to study ices in Class~II systems \citep{Pontoppidan2005,Pontoppidan2007,Terada2007,Schegerer2010,Terada2012a,Terada2012b,Terada2017}.
Since these sources require high sensitivity to be observed, due to their edge-on nature, JWST is the first observatory that allows for spatially resolved, sensitive observations of these systems with full wavelength coverage in the near- and mid-infrared.
An initial analysis of Near-Infrared Spectrograph \citep[NIRSpec][]{Jakobsen2022} data of the edge-on disk HH~48~NE detected spatially resolved ice features for the first time in a disk, including the trace species \ce{NH3}, OCS, \ce{OCN-}, and \ce{^13CO2}, and revealed a surprisingly large vertical extent of \ce{H2O}, \ce{CO2}, and CO ice absorption \citep{Sturm2023_NIRSpec}.
In this paper, we present new Mid-InfraRed Imager \citep[MIRI][]{Rieke2015} observations of the same source to complete the inventory of ice features observable with JWST.

Our radiative transfer models and interpretation of JWST NIRSpec observations have shown that the complex scattering light path through edge-on disks precludes a straightforward analysis of the ice features, so that ice abundance distributions can only be constrained by radiative transfer modeling of multi-wavelength observations \citep{PaperII,Sturm2023_NIRSpec}.
Ice feature profiles become distorted by scattering \citep{Dartois2024}, and their optical depths do not depend linearly on the column density along a direct line-of-sight toward the star because the multiple light paths through the disk partially `dilute' the absorption profiles.
Additional modeling diagnostics of the different ice absorbing regions are necessary in order to understand the spatial distribution of abundances in disks.

Another important element of protoplanetary disks are the Polycyclic Aromatic Hydrocarbons (PAHs). 
These big molecules (that in many aspects resemble small dust grains) play a crucial role in the charge balance, chemistry, and heating/cooling of the disk and could contain a significant fraction of the available carbon \citep{Tielens2008, Kamp2011}.
PAHs can be excited by UV radiation and cool efficiently through mid-infrared emission features. 
These features are attributed to the stretching of C--H (3.3~\micron) and C--C (6.23, 7.7~\micron) bonds, in-plane (8.4~\micron) and out-of-plane (11.27~\micron) C--H bending motions or a combination of both (5.4, 5.65~\micron).
The shape and comparative strength of the features can vary depending on the excitation state, charge, size distribution and hydrogenation \citep{Draine2007, Boersma2009, Geers2009, Maaskant2014}.
The distribution and excitation of PAHs in protoplanetary disks has been actively studied \citep{Habart2006, Visser2007, Geers2007, Bouteraon2019, Kokoulina2021}.
Modeling shows that K--M stars ($T_*<4200$~K) lack UV flux to excite the PAHs \citep{Geers2009}.
PAH features have been detected in numerous Herbig Ae/Be stars, but not in T~Tauri systems with a spectral type later than G8 \citep{Geers2006, Tielens2008, Maaskant2014, Lange2023}, except for the edge-on disk around Tau~042021 \citep{Arulanantham2024}. 
The authors suggest that the unique geometry of edge-on systems, which obstructs the thermal emission from the inner disk, enables the detection of the comparatively weaker PAH emission from T Tauri stars at mid-infrared wavelengths.

The paper is organized as follows: we first describe the details of the observations in Sect.~\ref{sec:observations} and the results that can be directly inferred from the observations in Sect.~\ref{sec:Observational results}. 
Then we describe the model setup for simulating observations of ices in protoplanetary disks in Sect.~\ref{sec:model setup} and use these models to interpret the ice composition, and abundance distribution of the HH~48~NE disk in Sect.~\ref{sec:modeling results}. 
Finally, we discuss the implications of our findings and the prospects of analyzing ices in edge-on protoplanetary disks in Sect.~\ref{sec:discussion}. We summarize our findings in Sect.~\ref{sec:conclusion}. 

\begin{figure}[!b]
    \centering
    \includegraphics[width = \linewidth]{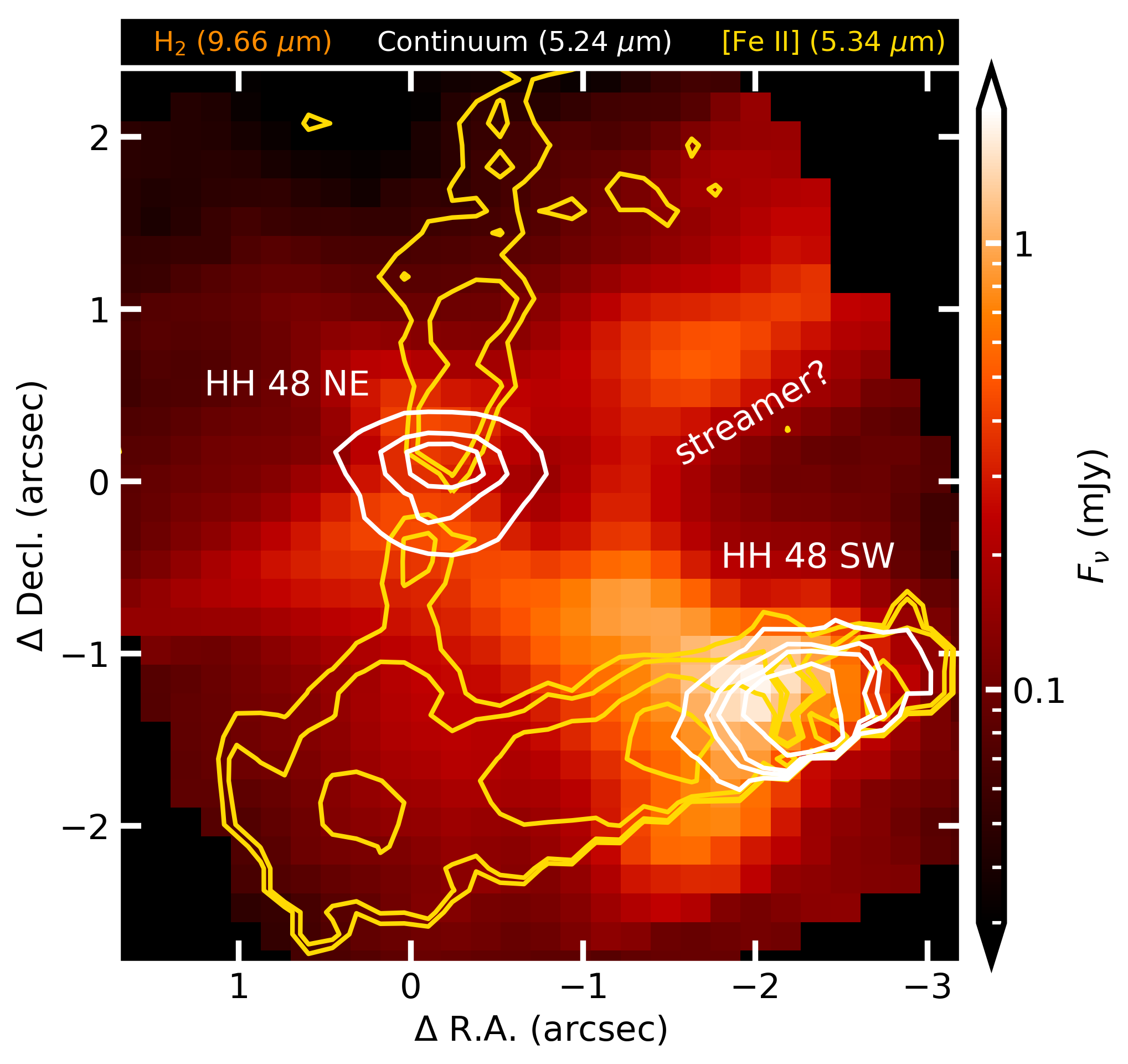}
    \caption{Overview of the HH~48 system at MIRI wavelengths. The continuum subtracted H$_2~V~=~0~-~0, J~=~5~-~3$ line emission (9.66~\micron) is shown in color tracing a complex structure of outflows and potential infall. The continuum at 5.24~\micron is shown in white contours at 20, 60, 100, and 300~\textmu Jy, tracing the scattered light from the two protoplanetary disks. The [Fe~II] line at 5.34~\micron is shown in the yellow contours at 5, 10, 30, and 100~\textmu Jy tracing the jets of both protostars. The primary, HH~48~SW and secondary, HH~48~NE are labeled in white.}
    \label{fig:2D gasoverview}
\end{figure}
\begin{figure*}[!b]
    \centering
    \includegraphics[width = \linewidth]{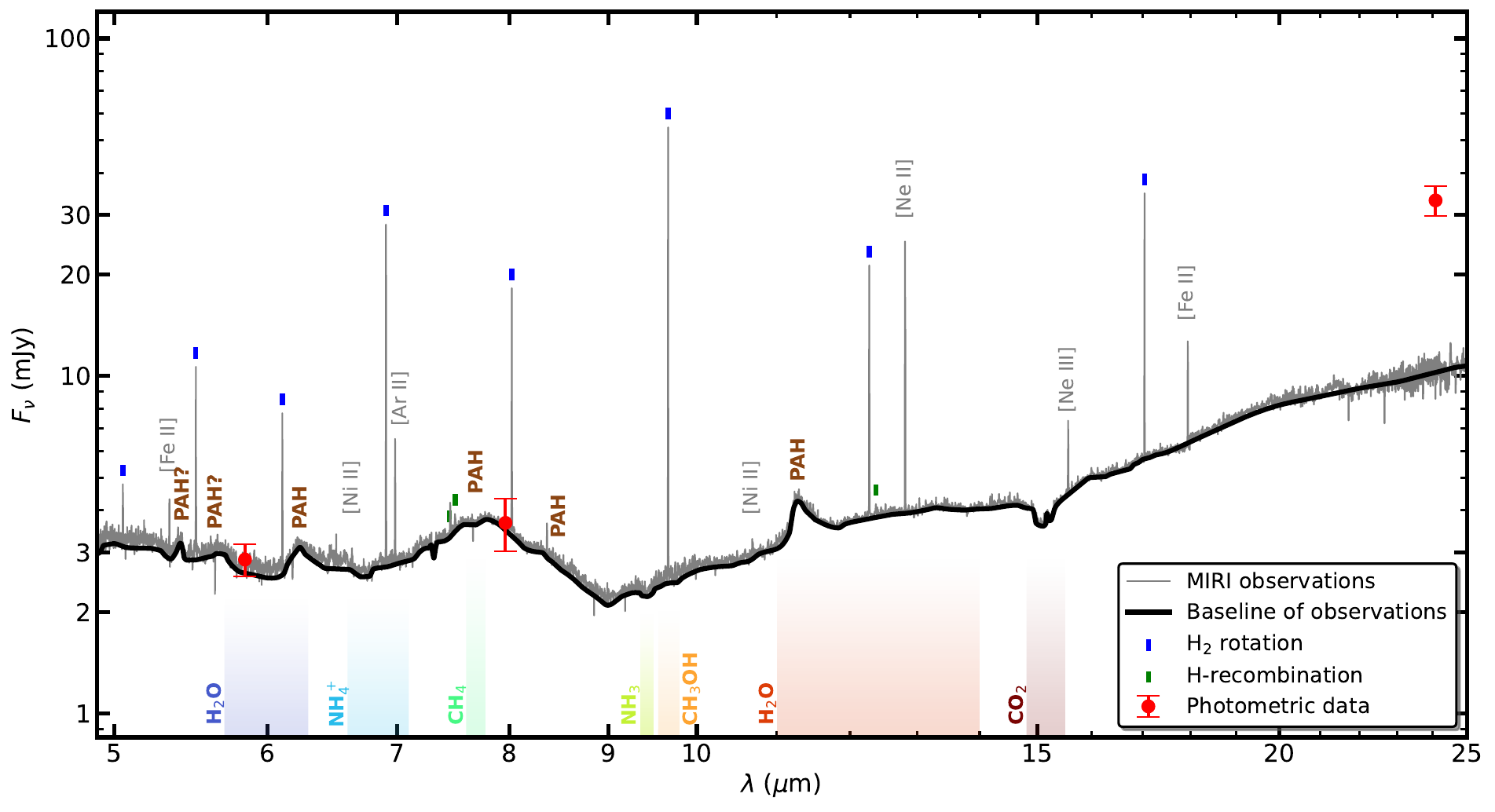}
    \caption{Integrated MIRI spectrum of HH~48~NE. The continuum baseline underneath the gas emission lines is shown as a black line. Gas features are labeled accordingly, molecular and recombination hydrogen lines are marked with colored tick marks. Absorption wavelength regions for important ices are marked on the bottom. \textit{Spitzer} photometric data points are shown in red.}
    \label{fig:overviewspectrum}
\end{figure*}
\section{Description of observations}\label{sec:observations}
We present new observations of the HH~48 system using MIRI/MRS onboard JWST, taken on 2023 March 23, as part of the Director's Discretionary Early Release Science (DD-ERS) program ``Ice Age: Chemical evolution of ices during star formation'' (ID 1309, PI: McClure). 
The spectroscopic data were taken with a standard 4-point dither pattern for a total integration of 1665~s, with the pointing centered on 11$^\mathrm{h}$04$^\mathrm{m}$23.18$^\mathrm{s}$, $-77$\degr18$^\prime$06.75$\dprime$. 
No target acquisition was taken because of the extended nature of the source.
A second, emission-free pointing was observed at 11$^\mathrm{h}$04$^\mathrm{m}$24.48$^\mathrm{s}$ $-77$\degr18$^\prime$38.77$\dprime$ with similar integration time to subtract the astrophysical background from the target flux. 
The data were processed through JWST pipeline version 1.11.4 \citep{bushouse2023_jwstmiri}, which includes the time-dependent correction for the throughput of channel 4.
The calibration reference file database versions 11.16.21 and jwst\_1119.pmap were used, which includes updated onboard flat-field and throughput calibrations for an absolute flux calibration accuracy estimate of 5.6 $\pm$ 0.7\% \citep{Argyriou2023}.  
The standard steps in the JWST pipeline were carried out to process the data from the 3D ramp format to the cosmic ray corrected slope image.  
The scientific background is subtracted after the ``Level 1'' run.
Further processing of the 2D slope image for assigning pixels to coordinates, flat fielding, and flux calibration was also done using standard steps in the ``Level 2'' data pipeline calwebb\_spec2.  
To build the calibrated 2D IFU slice images in the 3D datacube, the ``Stage~3'' pipeline was run. 
The final pipeline processed product presented here was built into 3D with the outlier bad pixel rejection step turned off, as this over-corrected and removed target flux. 

Each spectrum was extracted using a conical extraction, increasing the circular aperture with wavelength by employing a radius of 2.5 resolution elements and a minimum radius of 1\farcs{25}. 
Since the target is a binary, we extracted spectra from both sources at the same time, centered on 
11$^\mathrm{h}$04$^\mathrm{m}$23.1806$^\mathrm{s}$ $-77$\degr18$^\prime$04.744$\dprime$ (HH~48~NE) and
11$^\mathrm{h}$04$^\mathrm{m}$23.186$^\mathrm{s}$ $-77$\degr18$^\prime$06.939$\dprime$ (HH~48~SW).
Due to the increasing size of the point spread function (PSF) with wavelength, the sources spatially overlap at wavelengths $>12$~\micron.
To make sure that our target of interest, HH~48~NE, is the least contaminated by the binary component, HH~48~SW, the mask of HH~48~SW was used to deselect the contaminated region on the mask of HH~48~NE (see Fig.~\ref{fig:masks}).
The masks overlap at wavelengths >12~\micron, but the brightest core of HH~48~NE's disk remains resolved up to $\sim$25~\micron.
The 12 sub-channels are then scaled according to the median of the overlapping region in between the sub-channels starting from the shortest wavelengths (NIRSpec) to the longest sub-channels.
The correction is in all cases less than 5\%. For channel 4 ($\lambda$~>~18~\micron) we did not use any scaling because the difference in pixel size in combination with the double mask during extraction would result in sudden jumps in the spectrum.

\section{Observational results}\label{sec:Observational results}
An overview of the HH~48 system is presented in Fig.~\ref{fig:2D gasoverview} where we show the spatial extent of the continuum subtracted H$_2~V~=~0~-~0, J~=~5~-~3$ line with contours of the continuum at 5.24~\micron (white) and the continuum subtracted [Fe~II] line at 5.34~\micron (yellow).
HH~48 is a binary of two T Tauri stars with a separation of $\sim$2\farcs{3} or 425~au. 
Both sources show extended continuum emission at 5.24~\micron.
A jet is detected in both systems in the [Fe~II] line emission at 5.34~\micron.
The noble gas lines from Neon and Argon exhibit a spatial distribution very similar to that of the [Fe~II] lines, tracing the jet.
The \ce{H2} emission (background in Fig.~\ref{fig:2D gasoverview}) extends far beyond the scattered light continuum and probably traces disk winds in both protoplanetary disks.
Large-scale structure is detected in \ce{H2} toward the northwest of both sources that is not connected to any of the protostars in a coherent manner and traces the asymmetric structure earlier observed in CO~$J~=~2~-~1$ ALMA observations (see Fig.~\ref{fig:streamer} and \citealt{PaperI}). 
This component only appears in low-energy \ce{H2} lines ($E_{\rm up}<2000~\mathrm{K}$), which implies that the gas is cold with a high column density.
A careful investigation of the region did not show any signs of an additional source, which means that the most likely origin of the large-scale structure is an infalling streamer feeding the system or an interacting large-scale wind.
The focus of this paper is on HH~48~NE, as it is the more inclined of the two sources, but a spectrum of HH~48~SW is shown in Appendix~\ref{appendix: spectral extraction}.

\begin{figure*}
    \includegraphics[width = \linewidth]{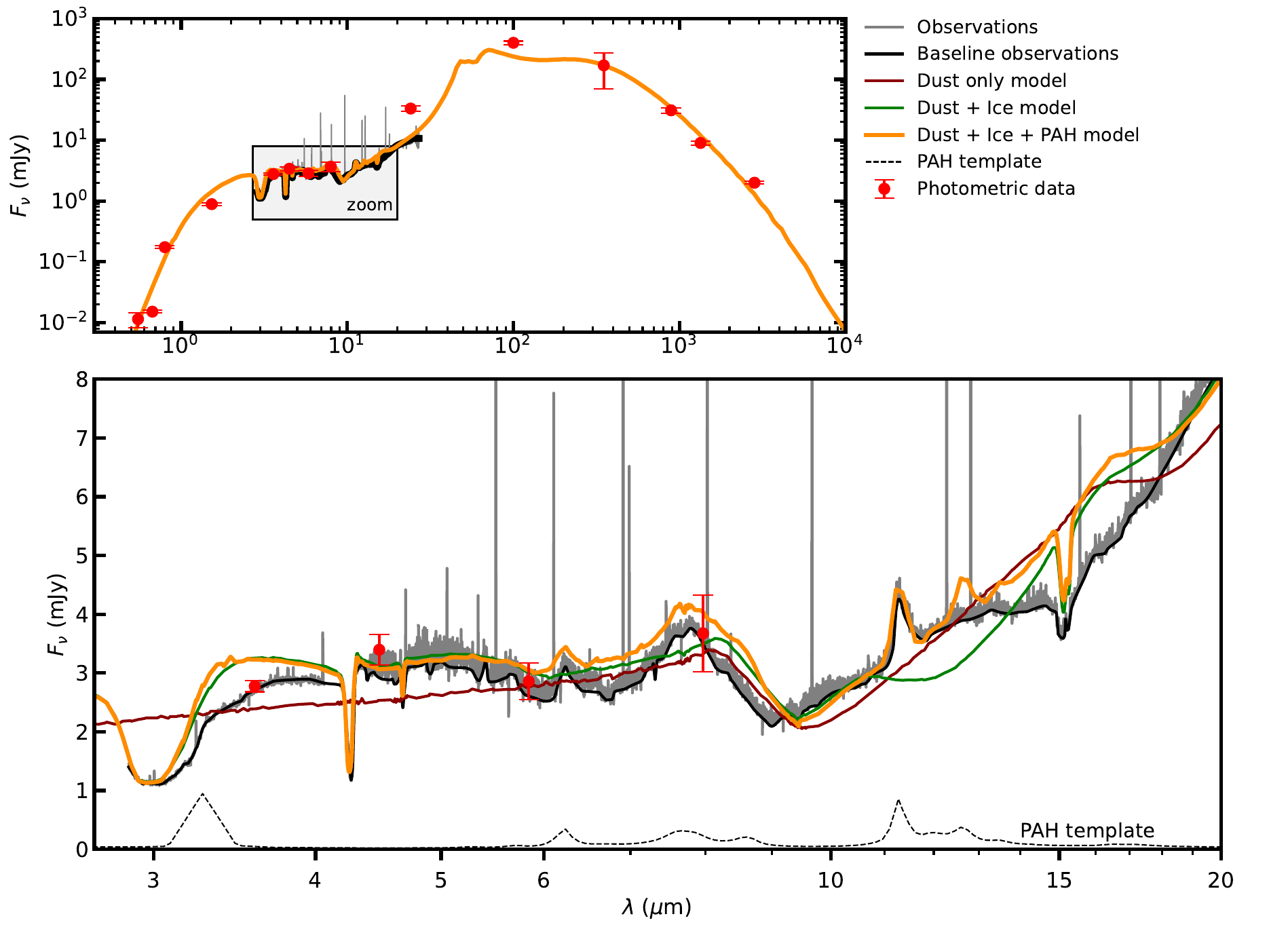}
    \caption{Overview of the source integrated spectrum of HH~48~NE (grey), with the continuum baseline underneath the gas lines shown in black with a zoom-in on the JWST wavelength range 2.7~--~27~\micron. Different ice radiative transfer models are shown for comparison with the observations. The initial model constrained from photometric data points, HST, and ALMA observations \citep{PaperI,PaperII} is shown in blue. A similar model was fine-tuned to the NIRSpec and MIRI spectrum with only refractory dust (red), dust and ice (green), and dust, ice, and PAHs (orange). The PAH opacity template used is shown in the bottom without extinction corrections \citep{Visser2007}.
    }
    \label{fig:SED comparison}
\end{figure*}

%\subsection{Integrated spectrum}\label{ssec: observational results - integrated spectrum}
Figure~\ref{fig:overviewspectrum} presents the source integrated MIRI spectrum of HH~48~NE.
Common gas emission lines are marked, molecular gas emission lines (e.g., CO, \ce{H2O}) can be seen throughout the spectrum but will be analyzed in a future article, as the main focus of this paper is on the ice absorption features.
The spectrum shows a shallow silicate feature in absorption that peaks at 9.0~\micron.
The peak of this feature is offset compared to the peak observed in emission in systems with lower inclination (usually around 10~\micron), due to an extra emission component from the inner disk that is scattered similarly to the continuum, as also predicted by radiative transfer models \citep{PaperII}. 
The observations are overall in good agreement with the photometric data points compiled in \citet{PaperI}, except for the \textit{Spitzer}/MIPS data point at 24~\micron. 
This photometric data point was based on a 2D Gaussian fit to the low-resolution \textit{Spitzer} data, which resulted in an overestimation by a factor of 3.

To estimate the continuum baseline underneath the molecular gas emission lines, we spline-fitted a line through the troughs between the lines including the broad features from PAH and ices in the continuum.
Since the placement of this continuum is uncertain, all figures include the full data as well for comparison.

\subsection{PAH emission}\label{ssec: observational results - PAH emission}
The most striking features are the PAH emission features most clearly seen at 6.23, 7.7, and 11.27~\micron, but also at 5.4, 5.65, and 8.47~\micron (see the PAH opacity template shown as the dashed line in Fig.~\ref{fig:SED comparison}, taken from \citealt{Visser2007}).
The PAH emission is spatially localized at the disk (i.e., we don't detect extended emission) and shows signs of severe wavelength dependent extinction since the 3~\micron feature is absent and the 6.2~\micron feature is weaker than expected. 
This suggests that the PAHs are located in the disk atmosphere and are obscured by its edge-on nature and/or that the ultraviolet (UV)/visible part of their excitation source is strongly attenuated.
Given the uncertainty of the extinction, it is difficult to compare feature ratios to constrain anything about the PAH characteristics.
The peak positions do not align with the established observational classes in ISM observations in \citet{Peeters2002} that usually correlate among different features; A potential combination of the three classes' characteristics is observed concurrently, indicative of radiative transfer spectral modification or active PAH chemistry in the disk. 
Although the 12.6~\micron feature is faint, it may be partially concealed by the \ce{H2O} libration band (as seen in the model without PAH in Fig.~\ref{fig:SED comparison}).
The 5.4 and 5.65~\micron features are more pronounced than anticipated compared to the 5.25 and 6.2~\micron bands, possibly due to the presence of small PAHs with multiple hydrogen groups in trios \citep{Boersma2009, Ricca2019}. 
Further exploration of the origin and characteristics of the PAHs is required to elucidate their significance in the physics and chemistry of regions where planets are forming, extending the analysis to less massive stars like T~Tauri stars.

\subsection{Ice features}\label{ssec: observational results - Ice features}
Strong ice absorption features, such as those observed in molecular clouds or protostars, are absent from our mid-infrared MIRI observations, which we interpret as a direct result of the multiple light paths that are possible through the disk \citep[see for more details][]{Sturm2023_NIRSpec}.
The strongest absorption feature is the \ce{CO2} bending mode at 15.5~\micron with a peak optical depth of only 0.2.
Due to the position of the PAH features and the effects of radiative transfer changing the profiles of the ice and silicate features, it is not appropriate to fit a polynomial to the continuum for ice identification and analysis.
Therefore, a radiative transfer model tailored to this source is required to discriminate ice absorption features from silicate and PAH features.

\subsection{Spatially resolved continuum emission in HH~48~NE}\label{ssec: observational results - spatial appearance}
\begin{figure*}
    \includegraphics[width = \linewidth]{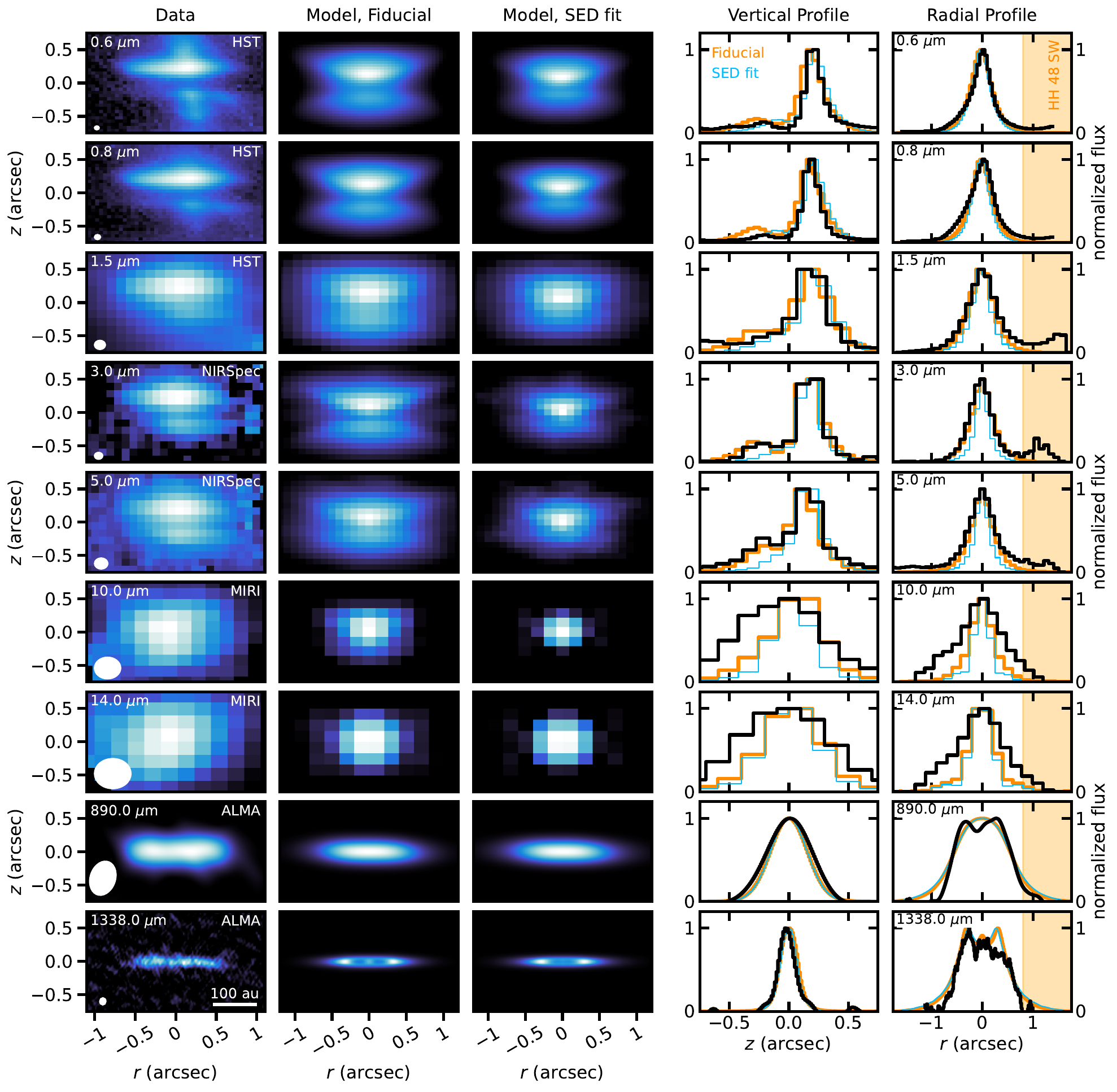}
    \caption{First column from the left: The spatial appearance of the continuum emission from HH~48~NE in the optical HST observations, JWST MIRI observations, and millimeter ALMA continuum observations. Second and third columns: spatial appearance of the continuum emission from the fiducial model and MIRI fit, respectively, after convolution to a representative PSF. Last two columns: median averaged vertical and radial profile of the continuum emission normalized to the peak (black) with the fiducial model (orange) and MIRI fit (blue) on top. The orange box marks the region where HH~48~SW starts dominating the continuum.}
    \label{fig:2D continuum comparison}
\end{figure*}

The left column of Fig.~\ref{fig:2D continuum comparison} presents the spatially-resolved continuum emission of the HH~48~NE disk at a range of wavelengths from optical HST observations to millimeter ALMA observations.
The bowl-shaped disk with a bright lobe on top, a dark lane, and a weaker lobe at the bottom, is recognizable up to 10~\micron. 
The disk is spatially resolved up to 15~\micron, which indicates that the continuum has a significant contribution from scattered light up to 15~\micron.
The East-West asymmetry that was observed previously in the optical HST observations is visible to a wavelength of $\sim$10~\micron.
Normalized vertical and radial cuts through the continuum emission are shown on the right.
Contributions due to HH~48~SW or possible dynamic interactions \citep{Stapelfeldt2014,PaperI} between the systems are masked in orange.

\section{Model setup}\label{sec:model setup}
\citet{Sturm2023_NIRSpec} show that the optical depth of the ice features in observations of edge-on protoplanetary disks is not linearly correlated with their column density and that the profiles of the ice features can be significantly altered due to the scattering opacity of the ices.
Since the disk is seen in scattered light in the JWST observations, it requires detailed 3D radiative transfer modeling to interpret the observed ice features and to constrain the region that is probed in the disk.

\begin{table*}[!b]
    \caption{Properties of the ice retrieved in the models compared to the ISM.}
    \begin{tabular}{L{.13\linewidth}|L{.05\linewidth}|L{.08\linewidth}L{.08\linewidth}|L{0.12\linewidth}L{.12\linewidth}L{.12\linewidth}L{.12\linewidth}}
        \bottomrule
        \toprule
        Ice species       & $E_\mathrm{b}$ (K)   & Disk $X$/H (ppm) & ISM $X$/H (ppm) & Disk $X/\ce{H2O}$ (\%)  &  ISM $X/\ce{H2O}$ (\%) & Comet $X/\ce{H2O}$ (\%) & 67P $X/\ce{H2O}$ (\%) \\
        \midrule
        \ce{H2O}                   & 5705 & 30   &  80    &  100  &   100   & 100         & 100  \\
        \ce{CO2}:CO 3:1            & 3196 & 6    &        &       &         &             &      \\
         -- \ce{CO2} fraction      &      & 4.5  &  22  &  15   &   13    & 2.0~--~30   & 4.7  \\    
         -- CO fraction            &      & 1.5  &  28    &  5  &   28    & 0.4~--~30   & 3.1  \\    
        \ce{CH4}                   & 1232 & 3.6  &  3.6   &  <12  &   1.9   & 0.4~--~1.6  & 0.34 \\
        \ce{NH3}                   & 5362 & 0.5  &  4.8   &  1.7  &   5.0   & 0.2~--~1.4  & 0.67 \\
        \ce{CH3OH}                 & 6621 & <0.5 &  4.8   &  <1.7 &   3.8   & 0.2~--~7.0  & 0.21 \\
        \midrule
    \end{tabular}
    \label{tab:ice properties}
    Only \ce{CO2} and CO are considered to be mixed in this work. The ISM ice abundances relative to hydrogen are taken from the inheritance model in \citet{Ballering2021}, the ISM ice abundances relative to water are taken from the background stars published in \citet{McClure2023}, cometary ice abundances relative to water are taken from \citet{Rubin2019}.
\end{table*}

\subsection{Physical structure}\label{ssec:model setup - physical and chemical structure}
We adopt the model setup described in \citet{PaperI,PaperII}.
The input stellar spectrum is assumed to be a black body with temperature ($T_{\rm s}$) scaled to the stellar luminosity ($L_{\rm s}$).
The density setup of the disk is parameterized with a power-law density structure and an exponential outer taper \citep{lynden-bell1974},
\begin{equation}
\label{eq:surface_density_profile}
\Sigma_{\mathrm{dust}}=\frac{\Sigma_{\mathrm{c}}}{\epsilon} \left(\frac{r}{R_{\mathrm{c}}}\right)^{-\gamma} \exp \left[-\left(\frac{r}{R_{\mathrm{c}}}\right)^{2-\gamma}\right],
\end{equation}
where $\Sigma_\mathrm{c}$ is the surface density at the characteristic radius, $R_\mathrm{c}$, $\gamma$ is the power-law index, and $\epsilon$ is the gas-to-dust ratio.
The inner radius of the disk is set to the dust sublimation radius, approximated by $r_{\rm subl} = 0.07\sqrt{L_{\rm s}/L_\odot}$, and the outer radius of the grid was set to 300 au (1\farcs{6}), which is consistent with the observations of the disk extent.

The vertical structure of the disk is described as
\begin{equation}
\label{eq:disk_height}
h=h_{\mathrm{c}}\left(\frac{r}{R_{\mathrm{c}}}\right)^{\psi},
\end{equation}
where $h$ is the aspect ratio of the gas and small dust grains, $h_{\mathrm{c}}$ is the aspect ratio at the characteristic radius, and $\psi$ is the flaring index.
We adopted two dust size populations to mimic dust settling in the system with a Gaussian distribution, 
\begin{equation}
    \label{eq:small_grains_distribution}
    \rho_{\mathrm{d}}=\frac{\Sigma_{\mathrm{dust}}}{\sqrt{2 \pi} r h} \exp \left[-\frac{1}{2}\left(\frac{\pi / 2-\theta}{h}\right)^{2}\right],
\end{equation}
where $\theta$ is the opening angle from the midplane as seen from the central star. 
The aspect ratio of the large grain population is restricted to $X\cdot h$ with $X \in [0,1]$.
The fraction of the total dust mass that resides in the large dust population was defined as $f_{\rm \ell}$.
The two dust populations followed a power-law size distribution with a fixed slope of $-3.5$ between the minimum grain size $a_{\rm min}$ and $a_{\rm max}$.
$a_{\rm min}$ and $a_{\rm max}$ of the small grains are left as free parameters.
The $a_{\rm min}$ value is consistent across both grain populations, while the $a_{\rm max}$ for the large grain population is fixed at 1~mm.
The dust and ice opacities were calculated using \texttt{OpTool} \citep{Dominik2021optool}, using the distribution of hollow spheres \citep[DHS;][]{Min2005DHS} approach to account for grain shape effects with the $f_\mathrm{max}$ = 0.8.
The gas is assumed to follow the small dust grain population, which implies that the gas-to-dust ratio varies over the vertical extent of the disk according to the dust settling.

The ALMA observations of HH~48~NE show clear signs of a central cavity, which is also supported by the spectral energy distribution (SED) and the HST scattered light observations \citep[See][for a detailed exploration of the source structure]{PaperI}.
Therefore, we incorporated a basic cavity description by scaling the dust surface density of both dust populations within the cavity radius ($R_{\rm cav}$) by depletion factor ($\delta_{\rm cav}$) - see Table \ref{tab:fid_mod_props}.

\subsection{Ice distribution}\label{ssec:model setup - ice distribution}
The ices are initially distributed at a constant abundance with respect to hydrogen within the expected ice surfaces, with the initial abundances taken from the inheritance model in \citet{Ballering2021}, based on observations summarized in \citet{Boogert2015}.
The model has an increasing number of ice species towards the midplane, based on their binding energy and a photodesorption limit.
The thermal desorption ice line of each system is determined using the approach described in \citet{Hasegawa1992}, that is the point where there is more ice than gas available of a given molecular species
\begin{equation}
    \frac{n_{\rm ice}}{n_{\rm gas}} = \frac{\pi a_{\rm d}^{2} n_{\rm d} S \sqrt{3k_{\rm B}T/M_{\rm n}}}{e^{-E_{\rm b}/T} \sqrt{2 k_{\rm B} N_{\rm ss} E_{\rm b}/(\pi^{2}m_{i})}} >1,
    \label{eq:gas-grain}
\end{equation}
where $n_{ice}$ is the number density of molecules in the ice phase, $n_{gas}$ is the number density of molecules in the gas phase, $a_{\rm d}$ is the characteristic dust grain size, $n_{\rm d}$ is the dust number density, $S$ is the sticking coefficient which is assumed to be 1, $k_{\rm B}$ is the Boltzmann constant, $T$ is the dust temperature, $M_{\rm n}$ is the molecular mass of the species, $E_{\rm b}$ is the binding energy, $N_{\rm ss}$ is the number of binding sites per surface area assumed to be $8 \times 10^{14}$~cm$^{-2}$ (\citealt{Visser2011}), $m_i$ is the mass of the species $i$. 
The region where ice is abundant is, in addition, vertically truncated by a limit that is initially put at $A_\mathrm{V}$ = 1.5~mag, motivated by the onset of water ice in dark cloud observations \citep{Boogert2015}, which corresponds to $z/r~\sim 0.3$.
Inside and above this limit, the model consists of bare grains without ice mantles.
This boundary is an approximation of the photodesorption limit and the slightly higher dust temperatures at low $A_\mathrm{V}$, which prevent ice formation on grains and assumes that all ice species are equally vulnerable to the same UV radiation.

In most parts of the disk, the vertical snowlines for \ce{H2O} and \ce{CO2} are determined by this photodesorption limit; the vertical snowline for \ce{CO2} is set by thermal desorption only within $\sim$20 au.
For species such as \ce{CH4} and pure CO, which have lower binding energies, the vertical snowline is exclusively determined by thermal desorption.

The average density of the dust and ice composite was then calculated in every region using
\begin{equation}
\rho_{\text {ice }}=\rho_{\rm gas} \frac{x_{\mathrm{ice}} M_{\mathrm{ice}}}{x_{\rm gas} M_{\rm gas}},
\end{equation}
where $x$ is the abundance with respect to total hydrogen, and $M$ the mean molecular weight. 
The gas predominantly consists of \ce{H2} and He and has an abundance of $x_{\rm gas} = 0.64$ and a mean molecular weight of 2.44.
The ice is distributed throughout the small and large grain population according to their surface area assuming smooth spheres, following the description in \citet{Ballering2021},
\begin{equation}
f_{\text {ice}, \ell}=\frac{f_{\text {dust}, \ell}}{\sqrt{a_{\max , \ell} / a_{\max , \mathrm{s}}}\left(1-f_{\mathrm{dust}, \ell}\right)+f_{\mathrm{dust}, \ell}}.
\end{equation}
In each of the two dust populations, the ice is distributed over the grains assuming a constant dust core - ice mantle mass ratio assuming efficient dust growth in the disk after initial freeze out in early phases of the star forming process. 

\begin{table}[!t]
    \centering
    \caption{Properties of the fiducial model presented in \citealt{PaperI,PaperII} and the model tweaked to fit the JWST spectrum.}
    \label{tab:fid_mod_props}
    \begin{tabular}{L{.3\linewidth}R{.33\linewidth}R{.22\linewidth}}
        \bottomrule
        \toprule
        parameter                       & fiducial                & best fit \\
        (unit)                          & \citep{PaperII}         & (this work)\\
        \midrule
        $L_{\rm s}$ (L$_\odot$)         & 0.41                    & 0.25\\
        $T_{\rm s}$ (K)                 & 4155                    & 4200\\
        $R_{\rm c}$ (au)                & 87                      & 87\\
        $h_{\rm c}$                     & 0.21                    & 0.24\\ 
        $\psi$                          & 0.13                    & 0.13\\ 
        $i$ $(^{\rm o})$                & 82.3                    & 83\\ 
        $\gamma$                        & 0.81                    & 1\\ 
        $M_{\rm gas}$ (M$_{\odot}$)     & $2.7\times10^{-3}$      & $3.2\times10^{-3}$\\
        $f_\ell$                        & 0.89                    & 0.97\\ 
        X                               & 0.2                     & 0.2\\ 
        $a_{\rm min}$ (\micron)         & 0.4                     & 0.3\\ 
        $a_{\rm max,s}$ (\micron)       & 7                       & 12.6\\ 
        $a_{\rm max,l}$ (\micron)       & 1000                    & 1000\\ 
        $R_{\rm cav}$ (au)              & 55                      & 50\\ 
        $\delta_{\rm cav}$              & $1.6\times10^{-2}$      & $3.5\times10^{-3}$\\
        photodesorption \qquad limit    & $A_\mathrm{V} = 1.5$ \qquad ($z/r \sim 0.3$)    &  $z/r = 0.6$ $(A_\mathrm{v}\sim0.1$) \\
        \bottomrule
    \end{tabular}
\end{table}

The additional mass of the ice is not added to the dust model, to keep the physical structure of the disk the same independent of the ice distribution. 
This implies that we effectively used a gas-to-solid ratio of 100 that includes silicates, amorphous carbon, and ice, instead of a gas-to-dust ratio of 100.
Depending on the choice of definition of settling (dust only, or dust and ice) and gas-to-dust ratio, the actual ice mass can vary up to a factor of a few, a point we return to in Sect.~\ref{sec:discussion}.

\subsection{PAHs}\label{ssec:model setup - PAHs}
The spectrum shows clear signatures of PAH emission, as described in Sect.~\ref{sec:Observational results}, and without properly considering its emission, it is difficult to identify and characterize the ice features. 
The PAH emission is spatially localized to the disk, and the relative strengths of the different features suggest that the PAHs are obscured, which means that they are emitting from within the protoplanetary disk.
Therefore, we model the PAHs in a simplistic way, including a population of PAHs in the model using the opacities from \citet{Visser2007} assuming a size distribution from 5 to 100~\AA\ and a power law slope of 3.5.
For the PAHs, we assumed the same temperature as for the dust; stochastic heating by UV photons was not taken into account. 
The PAHs are limited to the regions in the model where ices are absent due to either thermal desorption or photodesorption.
This assumption relies on the fact that PAHs are only excited in regions exposed to UV radiation and that their features are observed in emission rather than absorption.
The mass of PAHs required to match the strengths of the features in the observations is 175~ppm of the disk mass, or 17\% of the available carbon, which is consistent with ISM abundances and above previously modeled abundance that predict a depletion by orders of magnitude in Herbig Ae/Be stars \citep{Sloan2005,Visser2007,Geers2009,Lange2023}.

\section{Modeling results}\label{sec:modeling results}
As a starting point, we adopted the physical structure from \citet{PaperII}.
Initial ice abundances were established based on the average of their measurements relative to \ce{H2O} ice in \citet{Boogert2015}, with the \ce{H2O} abundance at $8\times10^{-5}$ following the reasoning in \citet{Ballering2021} and consistent with \citet{PaperII} (see Table~\ref{tab:ice properties}).
This model is a good fit to the spatial extent of optical HST scattered light observations, ALMA millimeter continuum observations, and the SED (see Fig.~\ref{fig:2D continuum comparison},\ref{fig:sed comparison sturm2023} and \citealt{PaperI}).
The model reproduces the observed JWST spectrum reasonably well, except for the silicate feature and the flux at wavelengths >20~\micron.
The main reason for this is the change of origin of the continuum emission as, in the models, the direct warm emission from the disk starts dominating over the scattered light of the warm inner disk in that particular region.
Since the initial model is already close to the data, we perturbed the parameters of the fiducial model using the MCMC setup presented in \citet{PaperI} with $L_\mathrm{s}$, $h_\mathrm{c}$, $\psi$, $M_{\rm gas}$, $f_\ell$, $a_{\rm min}$, $a_{\rm max,s}$, and $\delta_{\rm cav}$ as free parameters without waiting for conversion.
The best fitting model to the spectrum is given in Table~\ref{tab:fid_mod_props} with only minor changes in the setup.
Details on the outcome of the overall modeling and identified ices are presented in Sect~\ref{sec:ice identification}, and the effect of changing the size of the region in the disk where ices are included is investigated in Sect~\ref{sec:ice location}.

\begin{figure}[!t]
    \centering
    \includegraphics[width = \linewidth]{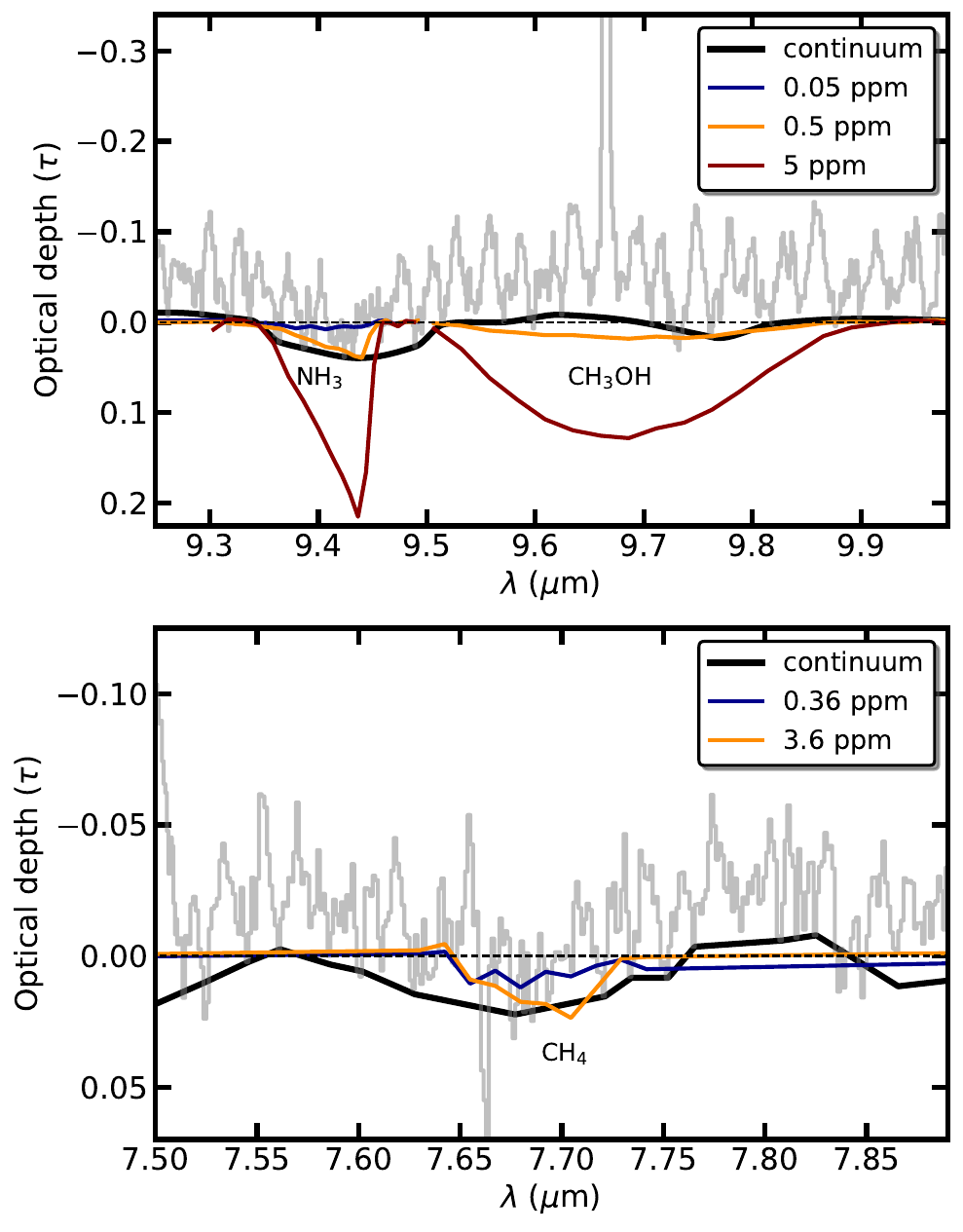}
    \caption{Optical depth representation of the \ce{NH3} and \ce{CH3OH} features (top) and the \ce{CH4} feature (bottom) with modeled features overlayed at different abundances. The black line indicates the continuum underneath the gas emission features including ices.}
    \label{fig:ch4 nh3}
\end{figure}
\begin{figure*}
    \includegraphics[width = \linewidth]{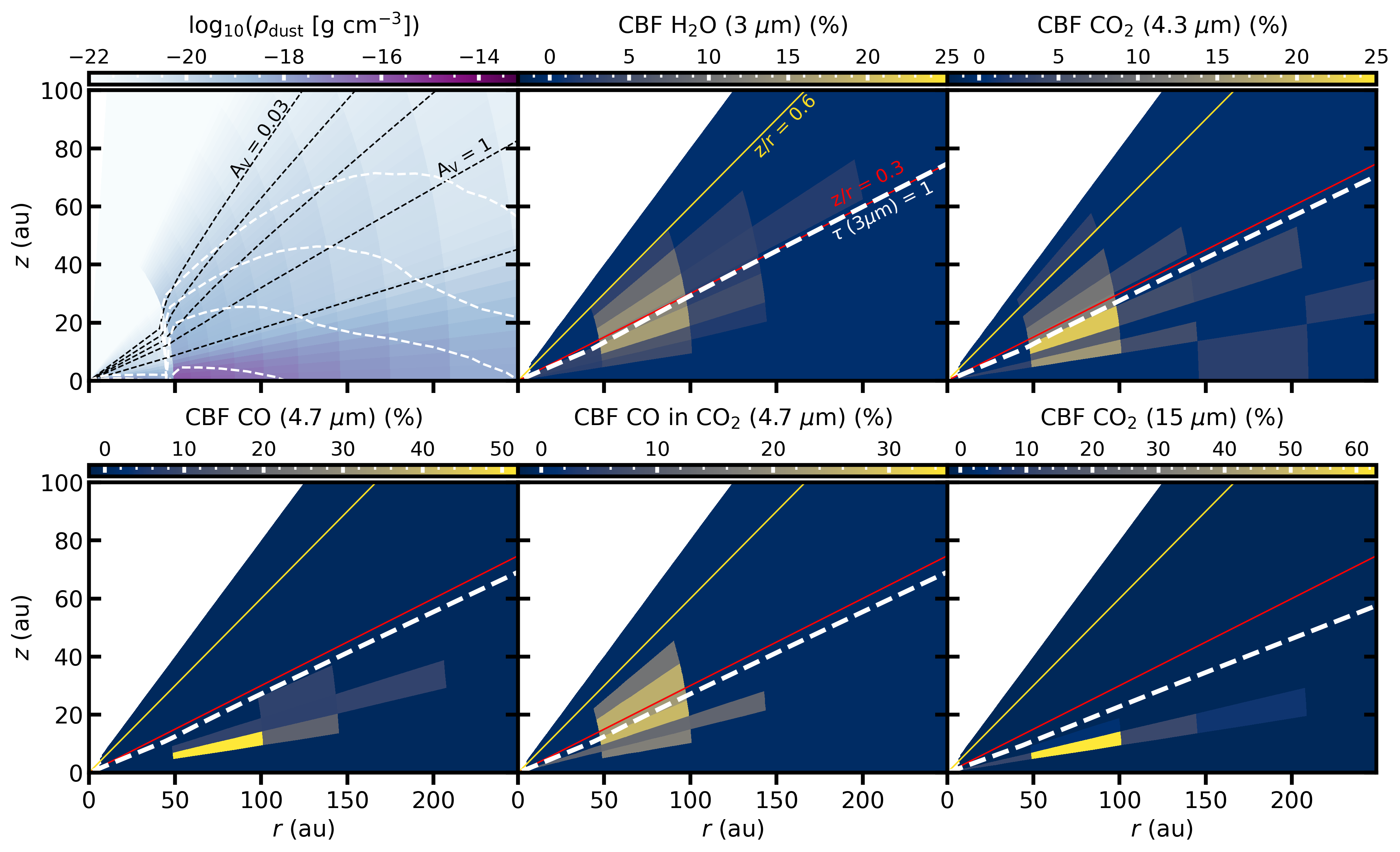}
    \caption{Top left: total dust density distribution in the model with internal shielding (radial $A_\mathrm{V}$; black dashed) at 0.03, 0.1, 0.3, 1, and 10 mag and external shielding (vertical $A_\mathrm{V}$; white dashed) at 0.01, 0.03, 0.1, and 1~mag from top to bottom. The other panels show the contribution function (CBF), i.e., the contribution of a specific region to the total absorption of the ice feature in percentages. The pure CO ice at high abundance is added for reference, but we argue in the main text that this component is not observed. The absorbing area is the region in the disk that contribute significantly to the total absorption. The red and yellow lines indicate the fiducial and best fit photodesorption snow surface, respectively, for reference. The thick, dashed, white line shows the $\tau=1$ surface at the ice feature wavelength. We would like to note that viewing angle is really important and is 83\degr\ in this case}
    \label{fig:CBF}
\end{figure*}

\subsection{Ice identification}\label{sec:ice identification}
In the NIRSpec spectrum (2.8~--~5.3~\micron), we identified the major ice components H$_2$O, CO$_2$, and CO, and multiple weaker signatures from less abundant ices NH$_3$, OCN$^-$, and OCS and the \ce{^13CO2} isotopologue.
Ice identification in the MIRI spectrum (5~--~28~\micron) requires comparison with a model to correctly predict the location of the ice features and to separate them from the PAH emission.
Fig.~\ref{fig:SED comparison} presents the comparison between the source-integrated spectrum of the observations and the best fitting model including dust, ice, and PAHs (orange line).
The model is overall a good representation of the spectrum with a few wavelength regions that are over-predicted.

\subsubsection{\ce{H2O}}
The bending mode of \ce{H2O} at 6~\micron is present in both the observations and the models, but is weak compared to the OH-stretch mode at 3~\micron.
Due to the PAH emission feature at 6.25~\micron it is hard to tell whether the discrepancy between the model and the observations is a result of the continuum being slightly too high or if the feature is stronger in the observations than in the model; other molecules could contribute to the absorption feature.
The libration mode of \ce{H2O} at 12~\micron is not observed in the spectrum, but the model without PAH emission (green vs. orange in Fig.~\ref{fig:SED comparison}) illustrates that the feature can also be hidden under the 11 and 12~\micron PAH emission features.

\subsubsection{\ce{CO2}}
The bending mode of \ce{CO2} at 15~\micron is weaker than expected from the \ce{CO2} stretch feature at 4.3~\micron (see Fig.~\ref{fig:sed comparison sturm2023}), and has the characteristic double-peaked profile of partially segregated \ce{CO2} ice \citep[see e.g.,][for a review]{Boogert2015}.
Partial segregation indicates that heating has induced diffusion, resulting in the formation of isolated pockets of pure \ce{CO2} ice within the matrix.
In addition, the bending mode has a broad contribution that is not accounted for in the model.
The best-fitting model includes one component of \ce{CO2}:CO at 20~K in a 3:1 ratio at an abundance of 6~ppm with respect to hydrogen.
This ice mixture was chosen to accommodate CO to be high up in the disk, and because it dominates the CO profile fit in \citet{Bergner2024}.
Using more mixtures could improve the fit to individual ice features in the MIRI range, but we refer to \citet{Bergner2024} for a detailed analysis of the ice profiles in the NIRSpec wavelength range.
The models show that we can roughly reproduce the strength and shape of the \ce{CO2} feature and the CO feature with this single component.
We revisit fitting the strength of both \ce{CO2} features in Sect.~\ref{ssec:ice location - photodesorption layer}.

\subsubsection{7~\micron region}
The 6.8~\micron absorption feature commonly found in protostars and dark clouds, usually attributed to \ce{NH4+}, is only weakly detected. 
Due to the PAH features at 6.25~\micron and 7~--~9~\micron, it is difficult to study this region in detail. 
However, the models suggest that there is an additional absorption feature centered at 6.8~\micron that is currently not taken into account in the model, which could be due to the \ce{NH4+} complex \citep{Keane2001,Boogert2008,Boogert2015}.
Ice absorption features of other species, for example \ce{SO2} and \ce{CH4}, are also known to occur at $\sim$7.6~--~7.7~\micron, the wavelength where the PAH feature is expected.
The PAH feature shows a dent at the peak, which can be reproduced with the addition of \ce{CH4} ice to the model at ISM abundance (3.6~ppm, \citealt{Ballering2021}) with respect to hydrogen, but a component including, for example \ce{SO2}, cannot be excluded (see Fig.~\ref{fig:overviewspectrum},\ref{fig:ch4 nh3}).
\ce{OCN-} has a feature in this range as well and is weakly detected at 4.6~\micron in the NIRSpec spectrum \citep{Sturm2023_NIRSpec}, but given the much weaker band strength at 7.6~\micron, the contribution is most likely negligible \citep{van_Broekhuizen2004}.
Since the shape of the PAH feature is relatively unconstrained, we leave a detailed analysis of this ice feature to a future project. 

\subsubsection{\ce{NH3} and \ce{CH3OH}}
Ammonia, \ce{NH3}, is detected at 9.3~\micron at a strength consistent with the feature found at 2.93~\micron (see Fig.~\ref{fig:ch4 nh3}). 
This feature is reproduced with a component of pure \ce{NH3} in the model with an abundance of 0.5~ppm with respect to hydrogen, a factor 10 lower than the abundance assumed in the ISM.
There is no clear sign of methanol, \ce{CH3OH}, in the spectrum: there is no sign of the 9.7~\micron \ce{CH3OH} feature and the red wing of the \ce{H2O} feature at 3.4~\micron shows no sign of the narrow 3.53~\micron \ce{CH3OH} feature \citep{Sturm2023_NIRSpec,TvS2018}, even though it cannot be excluded that there is a minor contribution (cf. the discussion in \citealt{Bergner2024}).
Given that the binding energy of methanol is higher than that of \ce{CO2}, this places tight constraints on the abundance in the disk regions that are traced with these observations. 
The maximum abundance of pure \ce{CH3OH} in the model that is consistent with the observations is 0.5~ppm.
We return to this in Sect.~\ref{sec:discussion}.

\begin{figure*}
    \includegraphics[width = \linewidth]{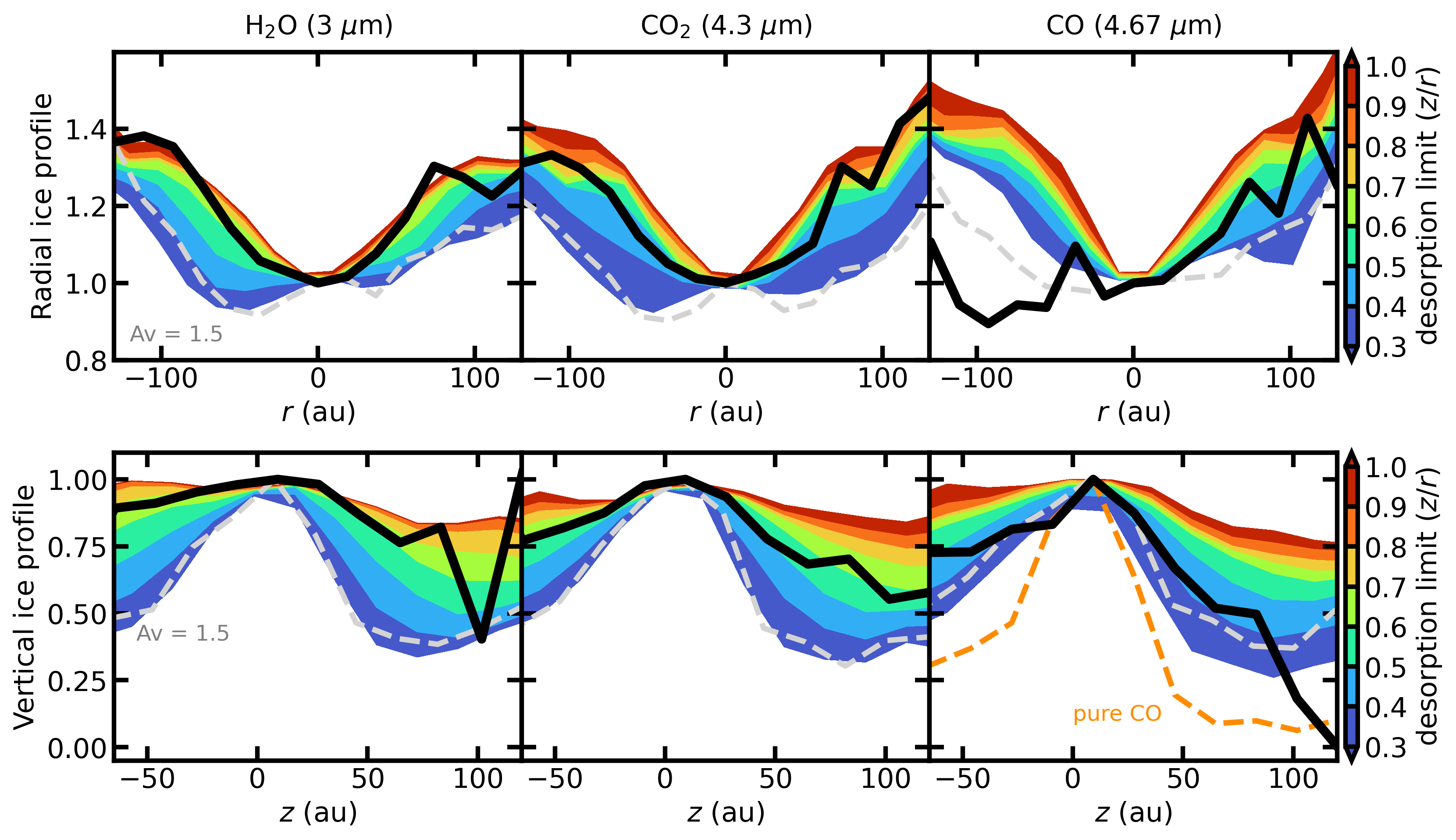}
    \caption{Median averaged profiles of the peak optical depths in radial direction (top) and vertical direction (bottom). The data are displayed in black and models with varying vertical photodesorption threshold in $z/r$ are shown in the colors specified in the colorbar. The gray dashed line shows the fiducial model with a photodesorption limit of $A_\mathrm{V} = 1.5$~mag and the orange dashed line indicates the vertical profile of the CO ice if pure CO ice is considered with a binding energy of 830~K.}
    \label{fig:absorption profiles}
\end{figure*}

\subsection{Localizing the measured ice}\label{sec:ice location}
Due to the complex radiative transfer between the emission and detection of the light, our observations do not trace the bulk ice reservoir in the midplane of the disk.
The presence of PAH emission complicates this further.
Furthermore, we have shown in \citet{Sturm2023_NIRSpec} that the spatially resolved information from the ice absorption is not directly related to the ice column density along the line of sight. 
The main reason for these results can be attributed to the position of the light source in our observations, which is situated at the center of the source rather than behind the disk as for background star observations.
Consequently, we examine the ice characteristics along the multiple trajectories of the photons that scatter within the disk's atmosphere, rather than along a narrow beam passing through the disk.
In this section, we explore what regions our observations trace in the models, and constrain the region where the observed ices are located in the disk.

\subsubsection{Contribution function}\label{ssec:ice location - contribution function}
To constrain the region probed by our observations, contribution functions were generated for the ice absorption in the model.
Due to scattering, it is not possible to deduce the regional contribution from a single model, since it requires storing the complete trajectory of every emitted photon, which is not practical.
Therefore, we rerun the same model multiple times, each time with the ices removed except for a specific region in the disk to test the contribution to the absorption in the source-integrated spectrum of that specific region.
The disk is divided into five radial bins with edges at 0, 50, 100, 150, 200, and 300 au and 11 vertical bins at $z/r$ of 0, 0.1, 0.15, 0.2, 0.25, 0.3, 0.4, 0.5, 0.6, 0.7, 0.8, and 0.9, which results in a total of 55 models.
The course gridding is necessary to have sufficient S/N on the ice feature in the source-integrated spectrum in a reasonable computation time.
Note that we neglected photodesorption in this case to show what the contribution of ices to the total absorption would be at all positions in the disk.
The regional contribution to the absorption of the different ice features in the source-integrated spectrum is displayed in Fig.~\ref{fig:CBF}.
We included a model with one component of CO mixed in \ce{CO2} as given in Table~\ref{tab:ice properties}, and a separate model with one component of pure CO with a binding energy of 830~K.

By comparing the contribution functions in Fig.~\ref{fig:CBF}, we can identify three different categories.
First, we see that the \ce{H2O} and \ce{CO2} features in the NIRSpec wavelength range (3 and 4.3~\micron, respectively) are dominated by a region between a radius of 50 and 100~au and significantly above the midplane ($z/r$ = 0.2~--~0.5).
The source of continuum at these wavelengths is the warm inner disk \citep{PaperII}, scattered through the atmospheric layers of the disk. 
Ice absorption is dominated by layers that are optically thin enough so that not all the light is absorbed but have enough ice mass to absorb a significant amount of light in the ice feature.
The disk midplane is too optically thick to let photons escape, and the disk atmosphere ($z/r \simeq 0.5$) and outer disk ($r > 100$~au) have too low ice mass to significantly contribute to the total absorption in the spectrum.
There is a significant contribution from below the $\tau=1$ layer at the absorption feature wavelength (white dashed line), which is a result of the fact that a large fraction of the photons encounter multiple scattering and could scatter downward (cf. \citealt{PaperII,Sturm2023_NIRSpec}). 

Second, pure CO ice absorption would be dominated by a region in the model close to the midplane because of its physical location.
Assuming a binding energy of 830~K \citep{Noble2012} the pure CO ice is restricted to $z/r < 0.15$, which is reached by only a small fraction of the light.
Due to the small fraction of light that reaches the CO ice-rich layer, the optical depths in the model cannot reproduce the data even for a very high CO abundance \citep[c.f.][]{PaperII}.
Instead, CO is mixed with \ce{CO2} or \ce{H2O} (see \citealt{Sturm2023_NIRSpec} and Sect.~\ref{sec:ice location}), leaving little room for pure CO \citep{Bergner2024}. 
CO mixed in the \ce{CO2} matrix shows similar results as the \ce{CO2} feature, which is no surprise given the small difference in feature wavelength.

\begin{figure}[ht]
    \centering
    \includegraphics[width = \linewidth]{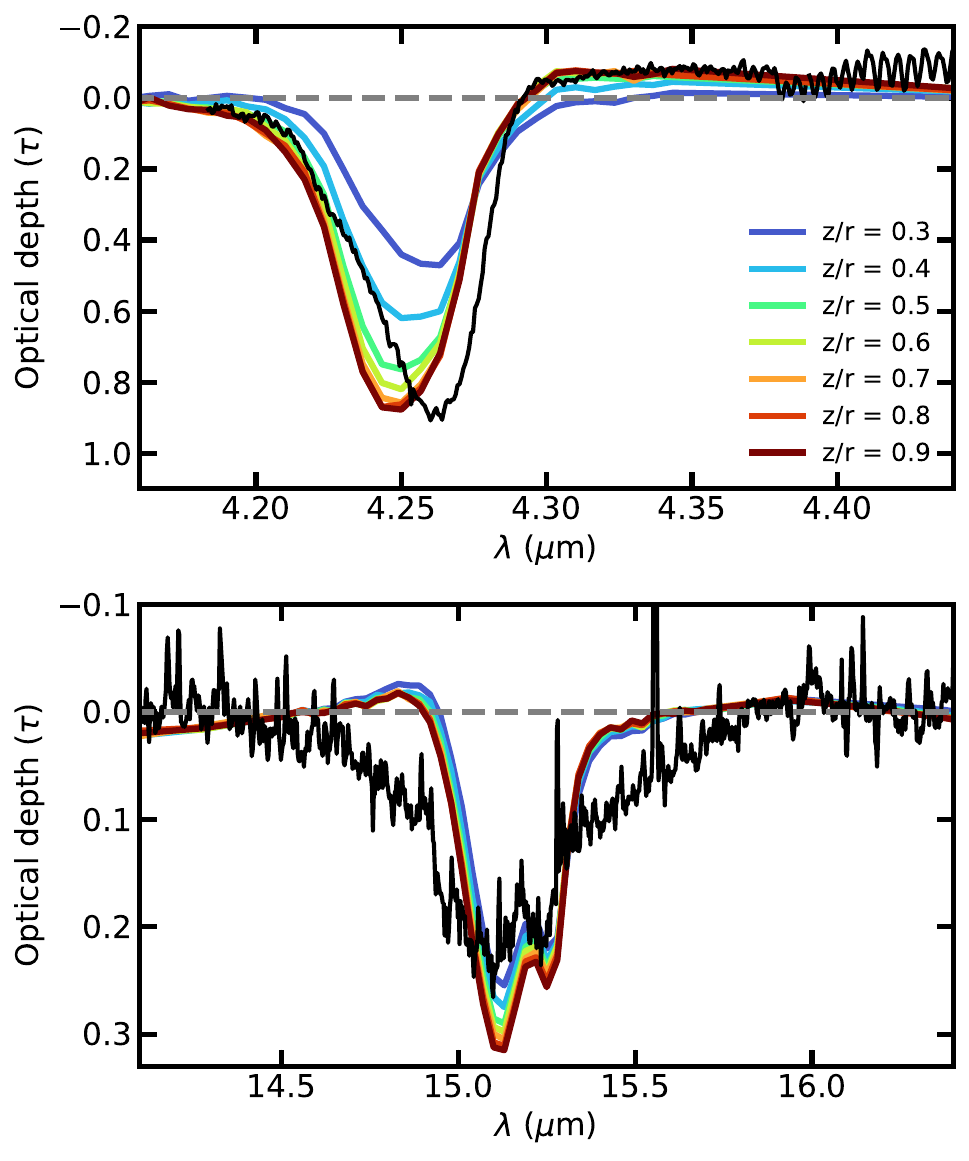}
    \caption{Comparison between the 4.3~\micron \ce{CO2} stretch feature and the 15~\micron bending mode in the observations (black) and models with different vertical photodesorption limits (colored lines). 
    The relative strengths of the features change with the photodesorption threshold due to a different spatial origin of the absorption (see Sect.~\ref{ssec:ice location - contribution function}). 
    The shape of the 4.3~\micron and especially the red emission wing at 4.30~--~4.35~\micron is also dependent on the height of the photodesorption limit due to geometric effects (see Fig.~\ref{fig:sketch}). No detailed fit to ice absorption profiles was performed, which could significantly improve the fit at 15~\micron \citep[see][for a fit of the 4.3~\micron feature]{Bergner2024}.}
    \label{fig:CO2_comparison}
\end{figure}

Third, the \ce{CO2} feature at 15~\micron traces a region close to the midplane between $z/r$ of 0.1 and 0.15.
The probed region with these features is closer to the midplane because of the longer wavelength of the features, which means that photons can penetrate further in the disk.
Additionally, the dominant source of continuum changes at these wavelengths from scattering to thermal dust emission in the models \citep[see][]{PaperII}, which means that the layer with the highest density that is still optically thin will produce the strongest absorption in the integrated spectrum.

\begin{figure*}
    \centering
    \includegraphics[trim = {3cm 7.5cm 1.5cm 0},clip,width = \linewidth]{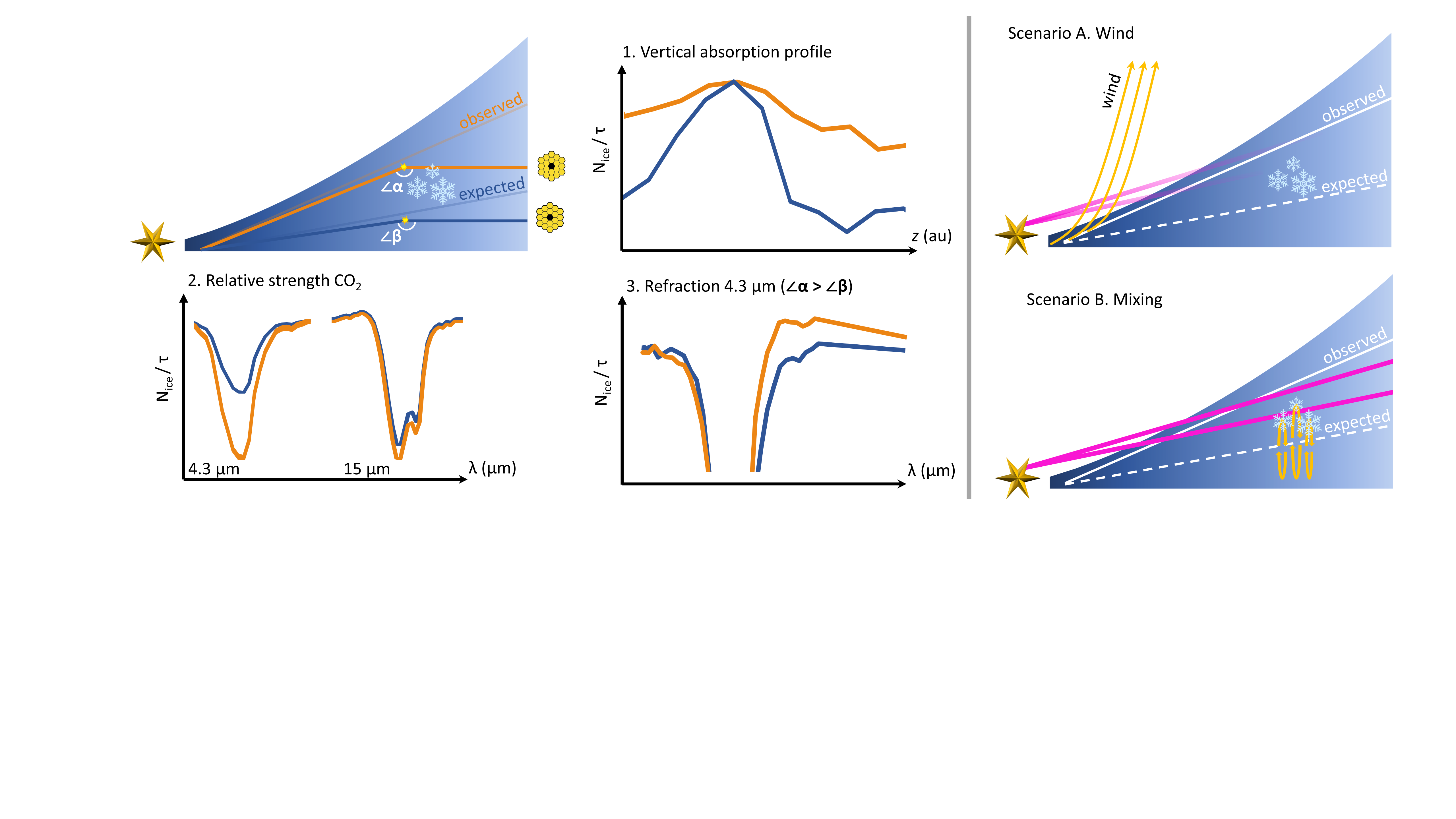}
    \caption{Sketch of the three pieces of evidence for elevated ices in HH~48~NE. The orange line demonstrates the effect of including the ices high up in the disk on the vertical absorption profile (panel 1.), the relative strength of the 4.3 to 15~\micron \ce{CO2} features (panel 2.), and the scattering wing of the 4.3~\micron \ce{CO2} feature (panel 3.). Two scenarios are proposed: a dusty disk wind blocking part of the UV emission from the star (Scenario A.) or vertical mixing stirring up ice to the surface (Scenario B.).}
    \label{fig:sketch}
\end{figure*}

\subsubsection{Locating elevated ices}\label{ssec:ice location - photodesorption layer}
There are three pieces of evidence indicating that the ices in the disk are elevated (located higher above the midplane) with respect to our model predictions assuming a photodesorption limit of $A\mathrm{_v}$ = 1.5~mag.
First, as discussed in \citet{Sturm2023_NIRSpec}, the vertical dependence on the absorption strength in the \ce{H2O}, \ce{CO2}, and CO features is less steep than expected if the ices were physically confined near the midplane. 
In Fig.~\ref{fig:absorption profiles}, we present the median averaged dependence of the peak optical depth as a function of both horizontal and vertical positions in the disk. 
These profiles are normalized to the absorption at $r~=~0$ and $z~=~0$. 
Unfortunately, the \ce{CO2} feature at 15~\micron is not spatially resolved, and the signal-to-noise ratio for the other ice features is insufficient for a comparable analysis. 
To determine the vertical extent of the ices, we conducted simulations with different photodesorption limits based on an ice destruction constraint in $z/r$. 
Subsequently, we convolved each wavelength point with the corresponding NIRSpec PSF using the \texttt{WebbPSF} package \citep{webbpsf2012} and calculated the median dependence of the optical depth in both horizontal and vertical directions similar to the observations. 
The modeled profiles are depicted in Fig.~\ref{fig:absorption profiles}. 
The gray line represents the standard model with a photodesorption limit of $A_\mathrm{V}$~=~1.5~mag, while the colored regions display the outcomes for different desorption limits. 
Even with the standard photodesorption limit, there are absorption features at high altitudes ($z~>~50$~au) caused by photons undergoing multiple scattering events, although this is insufficient to account for the observed profile. 
The best fitting model features a photodesorption threshold of $z/r$~=~0.6, corresponding to $A_\mathrm{V}$~$\sim$0.1~mag as observed from the star. 
The observed absorption profile for CO can only be explained by a component of CO mixed within a matrix with a higher desorption temperature, as pure CO is confined near the midplane due to its low desorption temperature, resulting in a sharp peak without a significant contribution from photons undergoing multiple scattering events ($z~>~50$~au). 

The radial profiles yield similar outcomes, although care has to be taken as these profiles are more sensitive to the source structure compared to the vertical profiles. 
One interpretation of our results would suggest that the \ce{CO2} feature aligns more closely with a photodesorption threshold of $z/r$~=~0.5, and the CO feature with a photodesorption threshold of $z/r$~=~0.4. 
However, given the uncertainties, we assume similar snow surfaces for all three molecules.

The second piece of evidence for elevated ices in the atmosphere of the disk is the weak absorption of the bending mode of \ce{CO2} at 15~\micron in comparison to the stretching mode at 4.3~\micron. 
The intensity of the ice absorption feature correlates with the extent of the disk area impacted by ice absorption.
The contribution function (shown in Fig.~\ref{fig:CBF}) indicates that the majority of the absorption at 4.3~\micron originates from a height between $z/r$ = 0.2~--~0.5, whereas the absorption at 15~\micron is predominantly from regions near the disk midplane ($z/r$ = 0.1~--~0.15). 
The standard model, with a photodesorption limit of $A_\mathrm{V}$ = 1.5~mag, successfully replicates the 15~\micron bending mode but fails to predict the \ce{CO2} stretch mode at 4.3~\micron by a factor of two (see Fig.~\ref{fig:CO2_comparison}). 
By expanding the region of absorption in the disk atmosphere by raising the photodesorption limit, the absorption at 4.3~\micron increases significantly more than that at 15~\micron, supporting a photodesorption limit at $z/r = 0.6$. 
Alternatively, increasing the abundance of \ce{CO2} in the model by about an order of magnitude in the $z/r$ = 0.2~--~0.5 region, as probed by the 4.3~\micron feature, could simultaneously account for both features with the fiducial photodesorption limit of $A_\mathrm{V}$ = 1.5~mag.
This would be consistent with some chemical models \citep[e.g.,][]{Drozdovskaya2016,Arabhavi2022}, but contradicts others \citep[e.g.,][]{Ballering2021}.

Finally, the shape of the ice features lends support to the notion of ice being located higher up in the disk because of geometric considerations. 
Apart from the importance of the extinction coefficient of ices, the refractive index of the ice also significantly influences the shape of the ice absorption features due to anisotropic scattering. 
In particular, additional absorption or emission wings are observed on either the blue or red side of the strong ice features \citep[see also][]{Dartois2022}. 
In the NIRSpec observations of HH~48~NE, apparent emission wings are detected on the red side and absorption on the blue side of both the \ce{CO2} and CO features at 4.3 and 4.67~\micron, respectively. 
This is opposite to the distortion observed in background star observations \citep[e.g.,][]{Dartois2022,McClure2023,Dartois2024} because of the source structure. 
For observations of background stars, the continuum is a consequence of direct emission from the star, and photons that scatter off the path on icy grains are not collected. 
However, for observations of disks viewed edge-on, the continuum is a result of scattered light, implying that photons that scatter from icy grains at a certain angle might be included in the observations.
These wings become more prominent when the photons scatter at extreme angles, indicating the presence of ice at higher elevations in the disk (see the sketch in Fig.~\ref{fig:sketch}). 
Fig.~\ref{fig:CO2_comparison} demonstrates that ice at $z/r \sim 0.6$ is required to explain the absorption at 4.18~--~4.21~\micron and apparent emission at 4.29~--~4.4~\micron.

\section{Discussion}\label{sec:discussion}
\begin{figure}[!t]
    \centering
    \includegraphics[width = \linewidth]{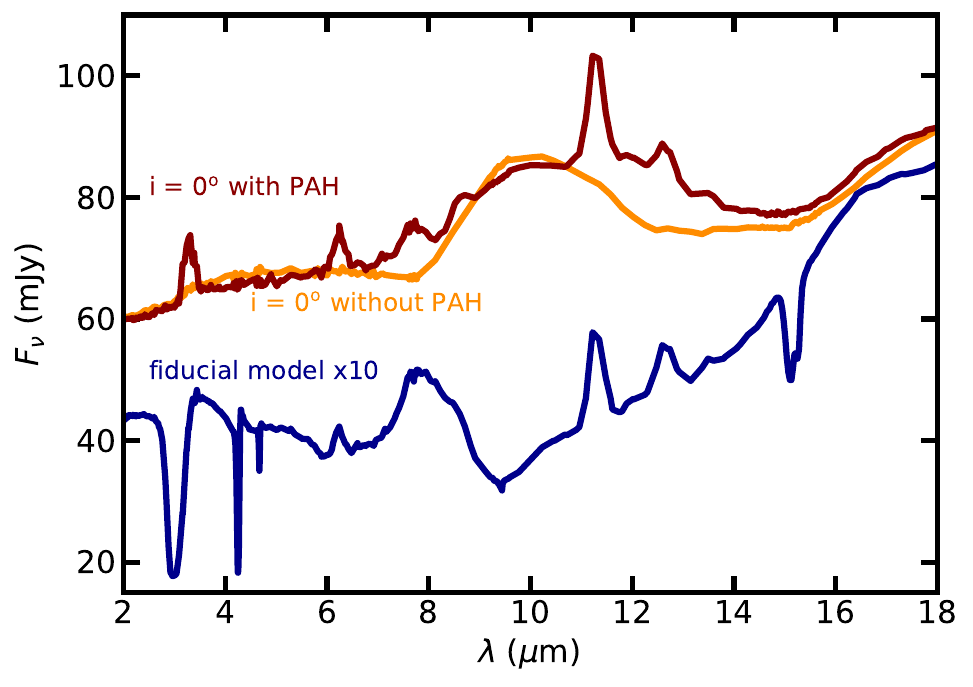}
    \caption{Comparison of the fiducial model (blue; $i=83$\degr) with face on disk models ($i=0$\degr) at similar opacities with (red) and without (orange) PAHs. The PAH emission is predicted to be strong in the face-on case as well, which is contradictory to literature observations of less inclined T~Tauri systems.}
    \label{fig:pah comparison}
\end{figure}

\subsection{PAHs in edge-on protoplanetary disks}\label{ssec:discussion pahs}
Protoplanetary disks that are observed face-on typically exhibit minimal or no indication of PAH emission, particularly within T Tauri star systems \citep[see e.g.,][]{Geers2006,Visser2007,Henning2024}. 
This absence is commonly understood to suggest a significant depletion of gas-phase PAHs by several orders of magnitude, combined with a lack of UV photons to excite them for stars with a spectral type later than G8 \citep{Geers2006,Tielens2008,Geers2009,Maaskant2014,Lange2023}.
We find mounting evidence that edge-on systems regularly show strong 8 and 11~\micron PAH features \citep[see also the spectrum of Tau~042021 in][]{Arulanantham2024}.
%Our radiative transfer model suggests that the relative strengths of the PAH features could be consistent with their abundances in the ISM. 

The fact that we see PAH emission in edge-on but not face-on disks suggests a localized origin of this emission. 
However, our best-fit model of HH~48~NE disk, if oriented face-on, still predicts strong PAH emission which is inconsistent with these previous observations (see Fig.~\ref{fig:pah comparison}).
The way in which we implemented PAHs in the model is simplistic and did not include stochastic visible/UV excitation, but we expect the resulting differences to be minor \citep{Woitke2016} and should not affect the conclusion that the PAH emission should be visible in face-on observations if they are equally distributed over the disk.
However, this implies that the PAH abundance estimated from our model should be considered an upper limit.
The inferred cavity in the HH~48~NE disk \citep{PaperI} might enhance the PAH emission due to reduced UV absorption by dust grains, thereby facilitating greater excitation of PAHs \citep{Geers2009,Maaskant2014}.
Additional modeling is required to explain this dichotomy in the observations between the two inclination classes.
Resolved observations of PAHs in edge-on and face-on systems would help constrain the properties of PAHs in disks \citep{Lange2023_resolvedpah}.

Bright PAH emission significantly complicates the analysis of ice features, as they dominate the continuum between 6 and 10~\micron. 
Therefore, we limit ourselves to the strongest ice features, and leave a search for the 6.8~\micron \ce{NH4+} feature and potential complex organic molecule signatures at 7.2 and 7.4~\micron \citep[see e.g.,][]{McClure2023,Rocha2024} to a future study.

\subsection{Ices at high elevation in protoplanetary disks}\label{Discussion: elevated ices}
\begin{figure}[!t]
    \includegraphics[width = \linewidth]{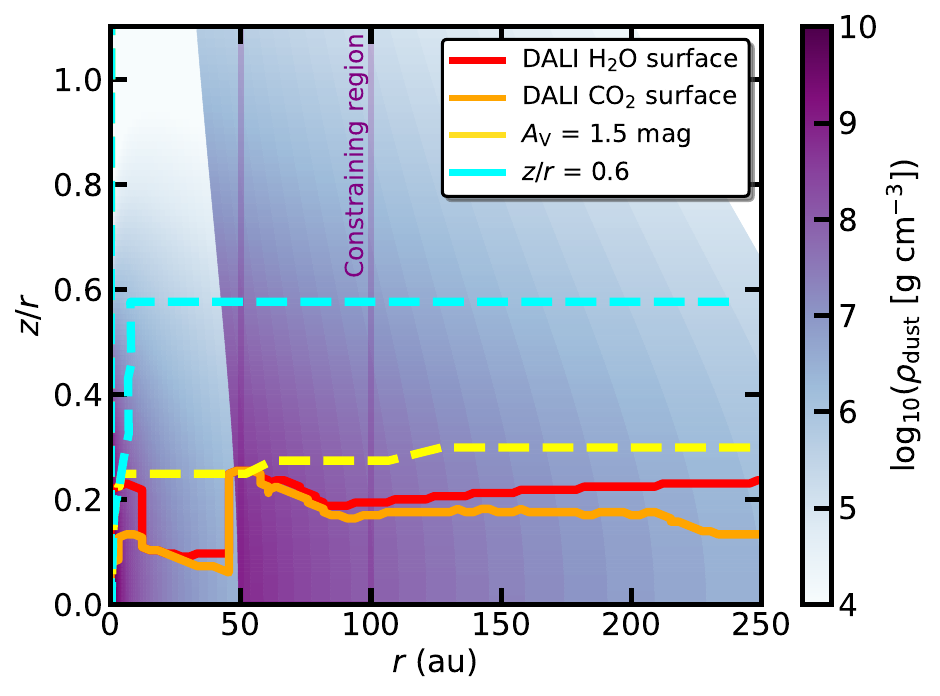}
    \caption{The snow surface as modeled with the simplified approach of $A_\mathrm{V}$ = 1.5~mag (yellow) and the required height (cyan). The other lines indicate the snow surfaces of \ce{H2O} (red) and \ce{CO2} (orange) in DALI models with a similar physical structure. The region that is predominantly probed by our ice observations (see Fig.~\ref{fig:CBF}) is marked with purple lines.}
    \label{fig:dalicomparison}
\end{figure}

We have shown using three different observables that at least part of the ice in HH~48~NE is located at a height of up to $z/r~=~0.6$.
The different methods are illustrated and explained in Fig.~\ref{fig:sketch}.
Our snow surface for the main ice species is well above the photodesorption threshold of $A_\mathrm{V}=1.5$, and is significantly above the thermal desorption threshold for \ce{CO}.
As an additional check for the position of the snow surface, we ran thermochemical DALI models \citep{Bruderer2012,Bruderer2013} with the same physical structure and dust opacities as in the RADMC-3D model.
This approach takes the full chemistry and molecular UV shielding (e.g. \ce{H2O} lines) into account and is therefore a better representation than our simplified approach assuming a photodesorption cut-off based on dust extinction.
We used the standard network that is based on the UMIST 06 network \citep{Woodall2007} to solve for steady-state chemistry.
Figure~\ref{fig:dalicomparison} presents the location of the snow surface, defined as the place where $X_\mathrm{ice} = X_\mathrm{gas}$, for the three main ice species.
The snow surface for \ce{H2O} and \ce{CO2} in DALI is consistent with our simple parametrization of $A_{\rm V} = 1.5$~mag, and agrees well in the region between 50 and 100~au which is the dominant region that is probed by our observations (see Fig.~\ref{fig:CBF}).

We consider two scenarios that could explain why the ices are elevated above a molecule's snow surface.
In the first scenario, part of the UV radiation is blocked, possibly by a dusty wind that shields the ices in the disk's atmosphere (see Fig.~\ref{fig:sketch}). 
Both a magnetically driven wind and a photoionized wind \citep{Olofsson2009,Owen2011,Panoglou2012,BoothR2021_dust_in_winds,Pascucci2023_PPVII} do have the potential to loft submicrometer-sized grains high into the atmosphere.
We do see evidence of \ce{H2} emission above the disk that likely originates in a disk wind, and found that the disk atmosphere is poor in sub-micrometer-sized dust grains, which could be because they are actively transported away from the atmosphere by the wind \citep{PaperI}.
Additionally, the system has an inner cavity which is depleted in dust and gas, which could be a result of internal photoevaporation. 
In the case of a magnetohydrodynamic (MHD) disk wind, models suggest that a high accretion rate of 10$^{-7}$~M$_\odot$yr$^{-1}$ is necessary to sufficiently block the disk from UV radiation \citep{Panoglou2012}.
Unfortunately, we do not have a reliable estimate of the mass accretion rate in the system because of the edge-on nature of the source, which blocks most of the emission coming from the inner disk.
However, considering the strong atomic jet and the potential streamer, it is not unlikely that the system is still strongly accreting.
It has been proposed before that a dusty disk wind might account for the elevated $A_\mathrm{V}$ values observed in a substantial sample of disks using the Ly alpha line, relative to the optical and infrared measurements \citep{McJunkin2014}. 
Further discussion on the implications and supporting evidence can be found in \citep{Pascucci2023_PPVII}.
UV shielding might also be increased by the apparently high concentration of PAHs in the warm regions of the disk (see Sect.~\ref{ssec:model setup - PAHs}), which effectively absorb UV radiation, or by molecular shielding from molecules like \ce{H2O} \citep{Bosman2022}.
If this scenario is true, it would have a significant impact on the interpretation of observations and models of protoplanetary disks, as UV radiation is in many cases a dominant driver of the chemistry \citep[e.g.,][]{Bergner2019_c2h_hcn,Bergner2021_Maps_HCN_CN,calahan_2022_uvchemistry} and can excite PAHs \citep{Visser2007,Geers2009}.

In an alternative scenario, the UV photons penetrate into the disk atmosphere as usual, but efficient mixing recycles the ices at higher altitude (see Fig.~\ref{fig:sketch}).
Chemical modeling including vertical turbulent mixing and diffusion shows that icy grains can reach the observable disk surface in turbulent disks, effectively mixing ices upward faster than they can be photodesorbed in the upper layers \citep{Semenov2006,Semenov2008,Furuya2013,Furuya2014,Woitke2022}.
The grain sizes of the icy grains inferred from the \ce{H2O} and \ce{CO2} features (>1\micron; see \citealt{Dartois2022} and \citealt{Sturm2023_NIRSpec}), which are thought to trace mainly the atmosphere of the disk, could be a direct result of mixing ice-rich micrometer-sized dust grains into the upper layers of the disk.
If this scenario is the case, we expect to see enhanced effects of UV-driven chemistry in the outer disk as a result of UV reaching a significant fraction of the ices in the disk \citep{Ciesla2012,Woitke2022}, which could be tested with (sub-)millimeter (e.g., ALMA or NOEMA) observations.

Additional analysis on a larger sample of sources is required to constrain which of these scenarios is most likely. 
Future directions could involve including a better description of photodesorption and PAH excitation using the stellar UV field and interstellar radiation field, and including molecular shielding for example by \ce{H2O} \citep{Bosman2022}.
Currently, external radiation is disregarded based on the assumption that sufficient residual cloud material shields the disk from such radiation; however, should it exert an influence, it would likely be counterproductive.

\subsection{Chemical implications}
The molecular abundances that we need to include in the model to match the observations are listed in Table~\ref{tab:ice properties}.
The retrieved abundances relative to hydrogen are considerably reduced, up to ten times less, when compared to earlier results reported in \citet{Sturm2023_NIRSpec} and \citet{Bergner2024}.
This is a direct result from the elevated snow surface which spreads the ices needed to reproduce the observed line depths over a larger disk volume compared to the fiducial photodesorption limit.
The presented abundances are a starting point for comparing observations with chemical models, including surface chemistry and vertical mixing.
Our absolute abundances are in general lower than ISM abundances taken from the inheritance model in \citet{Ballering2021}, this could be because we trace the ices in the disk atmosphere rather than in the midplane where the bulk of the ices reside.
Combining multiple features of the same molecular species would allow us to study the ice composition and abundance in different regions of the disk; however, as shown in Sect.~\ref{ssec:ice location - photodesorption layer} knowing the physical location of the snow surface beforehand is crucial as these properties are degenerate.
Comparing the height of the snow surface derived with JWST with high resolution ALMA observations (for example of CO, \ce{H2CO}, \citealt{Podio2020}) could help in this regard, but these are currently not available for this system.
If sequestered ice is responsible for the gas-phase molecular depletion of CO and \ce{H2O} \citep[see e.g.,][]{Krijt2020,Sturm2022CI}, then it is located in the midplane, where we cannot trace it with edge-on disk observations.

In Table~\ref{tab:ice properties} we compare the ice ratios with respect to water in the protoplanetary disk with the background star observations in \citet{McClure2023}.
The latter observations probe the ices in the molecular cloud stage in absorption against the continuum of a background star.
Since the HH~48~NE observations are taken not far from the molecular cloud, the initial conditions of the disk were likely very similar to the background star observations.
The overall \ce{CO2}/\ce{H2O} column density ratio is consistent with the background star observations, although the individual mixtures have changed \citep{Bergner2024}.
CO appears significantly less abundant in the protoplanetary disk, compared to background star observations.
However, it is very likely that the disk observations only probe the CO that is trapped in the \ce{CO2} or \ce{H2O} matrix and therefore survives high up in the disk, while the bulk of the CO ice is closer to the midplane where it cannot be probed by edge-on disk observations.
\ce{CH4} is tentatively detected at the location of a PAH emission features, which makes the observed optical depth uncertain.
Combined with the low binding energy, which indicates that the majority of the ice is located close to the midplane, the observed abundance is unreliable. 
\ce{CH4} is expected to be abundant in protoplanetary disks due to the active carbon chemistry \citep{Bruderer2013,Woitke2016,Krijt2020,Oberg2023_chemistry_review}, which is in line with the observations, but would be an order of magnitude higher than observed in comets and the ISM (see Table~\ref{tab:ice properties}).
Both \ce{NH3} and \ce{CH3OH} are significantly depleted ($\times 3$ and >$\times 3$, respectively) in the disk with respect to the dark cloud \citep{McClure2023}. 
Both of these molecules are not efficiently formed in protoplanetary disks, but predicted to be inherited predominantly from the natal cloud \citep{Drozdovskaya2016,Ballering2021,Booth2021_methanol}.
Their low abundance suggests that at least part of the ice is reset and/or processed upon entering the protoplanetary disk.
Alternatively, inherited ice ends up mostly in the midplane, and the ice we observe is formed locally in the disk.
The low abundance of \ce{NH3} and \ce{CH3OH} with respect to water is consistent with the observed abundances in comets. 
Presumably, comets are formed in the midplane, so in the midplane the \ce{CH3OH} abundance should be low as well, and thus favoring some reset.
We would like to note, however, that comets trace the history of the protosun which may not have been formed in an environment like HH~48~NE, so one-to-one comparison is non-trivial and will require ice observations in a larger sample of disks around solar-like stars.

\subsection{Limitations and future outlook}\label{limitations and future outlook}
We have shown that we can approximately reproduce the NIRSpec and MIRI spectrum with a relatively simple parameterized model including dust, PAHs, and ices.
In this section we will discuss the main observables that can be used to constrain ice properties from the observations and give a future outlook to the possibilities.

\subsubsection{Ice observables}\label{ssec: ice observables}
Due to the complex radiative transfer in edge-on disk observations, constraining properties of the ices is not trivial. 
Directly using the optical depth of the ice features to derive column densities can lead to errors of several orders of magnitude, which should be avoided \citep{Sturm2023_NIRSpec}.
There are a few observables that can be used to compare with models to learn more about the source structure, ice abundance and composition from the ice features.

\begin{enumerate}
    \item The strength of ice absorption features is related to the size of the absorbing region in the disk and the ice abundance. 
    \item The vertical distribution of the peak optical depths of the ice can be used to locate ice vertically in the disk. 
    In this work, we have shown that the ices in HH~48~NE are limited vertically by photodesorption, but a similar analysis can be done on thermal snow surfaces if, for example, pure CO ice is detected.
    Comparing the vertical dependence of the absorption in the 15~\micron \ce{CO2} feature with that of the 4.3~\micron feature will give additional constraints on the source structure, as this will allow to probe the extent of the absorbing area at this wavelength (see Fig.~\ref{fig:CBF}).
    \item Modeling observations of multiple features from the same species (e.g., \ce{H2O} and \ce{CO2}) will allow constraining the abundance and composition in different regions in the disk. 
    Moreover, their comparative intensities can serve as a gauge for determining the extent of the area that absorbs.
    \item The isotopic ratios (e.g. \ce{^13CO2}, \ce{^13CO}) with the main isotopologue give information on the level of saturation in the main isotopologue, assuming a \ce{^12C}/\ce{^13C} value.
    We have shown that our models can explain the strong \ce{^13CO2} absorption, but other isotopologues were unfortunately not detected.
    If, for example, \ce{^13CO} had been detected, this would hint at a small absorbing area with a high abundance of CO instead of being mixed in \ce{CO2} and therefore widespread in the disk.
    \item Ice profile shapes can be used to constrain the chemical environment of the ice which is done in a separate paper \citep{Bergner2024}.
    Combining multiple features can give insights into the vertical segregation of ice mixtures.
    This is only possible with high spectral resolution mid infrared instruments like MIRI.
    \item We have shown that scattering wings can be used to put constraints on the source structure and the scale height of the ice, taking the geometry of the system into account. Additionally, these features can be used to constrain the grain size distribution \citep[cf.][]{Dartois2022,Dartois2024}.
\end{enumerate}
In the end, a combination of these observables will result in the best constrained source structure in the model and the best constraints on ice properties.

\subsubsection{Future prospects}\label{future prospects}
In this study, we have demonstrated the feasibility of fitting JWST observations using a relatively straightforward approach.
All relevant observables discussed in Sect.\ref{ssec: ice observables} are simultaneously matched in the model.
Expanding the sample size is the next logical step to compare our results for HH~48~NE with other disks; currently, only this particular source has been thoroughly examined, and it may be an outlier due to its proximity to the companion HH~48~SW.
The methodology employed presents a practical means to model a broader array of disks and gather data on the composition and distribution of ice in protoplanetary disks. 
If other sources validate that the observed ices share a similar snowsurface as a result of photodesorption and ice mixtures, a reasonable simplification in the fitting process is a model featuring only three dust populations: small dust particles without ice, small dust particles with ice, and large dust particles with ice. 
This streamlines computational efforts significantly and allows for further adjustments or the inclusion of ices in MCMC or $\chi^2$ fitting. 

One drawback of using edge-on disks to investigate ices is that it does not capture the bulk of the ice mass in the disk midplane, but rather focuses on regions in the disk's atmosphere (see Fig.~\ref{fig:CBF}). 
Disks with lower masses have the potential to reveal a higher proportion of their ice content, with features being less likely to be saturated. 
Nevertheless, observed edge-on disks typically possess substantial mass by nature \citep{Angelo2023}, which complicates the analysis of ices. 

If it is indeed the case that the contrast of the PAHs in the disk is intensified due to the high inclination of this source, identifying subtle features will prove to be quite challenging. 
A more sophisticated incorporation of PAHs, including stochastic heating, is essential to make any assertions regarding their abundance and distribution within the disk. Furthermore, future endeavors should focus on comparing the results with ALMA data to differentiate between various scenarios. 
The scarcity of high-resolution ALMA observations of molecules in edge-on disks necessitates further exploration to impose additional constraints on the physical disk structure and chemistry.

\section{Conclusion}\label{sec:conclusion}
We presented new JWST/MIRI data to complete the ice inventory of the edge-on protoplanetary disk HH~48~NE. 
Radiative transfer models are fitted to the data to constrain abundances and the regions that are probed with the observations.
We therefore conclude the following.
\begin{itemize}
    \item Edge-on disks are promising laboratories for studying ices in protoplanetary disks, as we find clear evidence for various newly-detected disk ice features including \ce{CO2}, \ce{NH3} and tentatively \ce{CH4}. JWST spectra can be fitted well with 3D radiative transfer models using a relatively simple setup, which enables detailed analysis of their abundance and physical location in the disk.
    \item Ices are located unexpectedly high up in the disk, up to $z/r~\sim~0.6$ and $A_\mathrm{V}~\sim0.1$. This result is determined based on the vertical dependence of the absorption, the ratio of the 4.3 and 15~\micron \ce{CO2} features, and the scattering wings of the 4.3~\micron feature. We suggest two possible scenarios: (1) Entrained dust in a disk wind is blocking the stellar UV radiation, so ice may survive higher up in the disk, but this would have strong implications for the gas-phase chemistry as well. (2) Alternatively, the icy material is efficiently elevated in the disk by vertical mixing, continuously replenishing the ice reservoir in the upper disk atmosphere.
    \item JWST NIRSpec probes the ices in a higher layer in the disk than MIRI due to a change in scattering properties and the origin of the continuum. This complicates the analysis since the abundance and mixture of ices may change spatially in the disk, but with sufficient wavelength coverage, we can study multiple disk regions simultaneously.
    \item The absence of a methanol feature and the faint \ce{NH3} feature indicates that some of the ice from the original cloud is destroyed during earlier phases, reset upon entering the protoplanetary disk or remains concealed in the optically thick midplane.
    \item The PAH emission contrast to the continuum is strong in HH~48~NE, but is extincted by the disk, which implies that it must be localized to the disk itself. Current models cannot explain why the contrast in edge-on systems is so much different than in less inclined systems. This suggests that PAHs can be more abundant than previously thought, but only in specific regions in the disk probed by these observations. The spectrum of PAH emission bands hinders the analysis of ice features due to its strong, widespread impact on the continuum location.
\end{itemize}

JWST provides a unique opportunity to study ices and PAHs in protoplanetary disks in unprecedented detail.
Full NIRSpec and MIRI coverage is required to understand and constrain the full ice inventory in protoplanetary disks. 
The application of the analysis presented here to additional protoplanetary disks will allow a more global understanding of the composition and distribution of ices in these regions, which is important in the context of the evolution of planetary systems.

\begin{acknowledgements}
We thank the anonymous referee for the constructive feedback on the manuscript.
Astrochemistry in Leiden is supported by the Netherlands Research School for Astronomy (NOVA), by funding from the European Research Council (ERC) under the European Union’s Horizon 2020 research and innovation programme (grant agreement No. 101019751 MOLDISK).
M.K.M. acknowledges financial support from the Dutch Research Council (NWO; grant VI.Veni.192.241).
Support for C.J.L. was provided by NASA through the NASA Hubble Fellowship grant No. HST-HF2-51535.001-A awarded by the Space Telescope Science Institute, which is operated by the Association of Universities for Research in Astronomy, Inc., for NASA, under contract NAS5-26555.
D.H. is supported by Center for Informatics and Computation in Astronomy (CICA) grant and grant number 110J0353I9 from the Ministry of Education of Taiwan. 
D.H. acknowledges support from the National Technology and Science Council of Taiwan through grant number 111B3005191.
E.D. and J.A.N. acknowledge support from French Programme National ‘Physique et Chimie du Milieu Interstellaire’ (PCMI) of the CNRS/INSU with the INC/INP, co-funded by the CEA and the CNES. 
M.N.D. acknowledges the Holcim
Foundation Stipend.
Part of this research was carried out at the Jet Propulsion Laboratory, California Institute of Technology, under a contract with the National Aeronautics and Space Administration (80NM0018D0004).
S.I. and E.F.vD. acknowledge support from the Danish National Research Foundation through the Center of Excellence “InterCat” (Grant agreement no.: DNRF150).
M.A.C. was funded by NASA’s Fundamental Laboratory Research work package and NSF grant AST-2009253.
\end{acknowledgements}

\bibliographystyle{aa}
\bibliography{refs}

\begin{appendix}
\section{Streamer}
The \ce{H2} emission shows, apart from the disk winds in both systems, an additional emission component to the North-West that is not correlated with the orientation of any of the sources' outflow. 
A comparison with the Moment 0 map of the CO~$J~=~2~-~1$ emission presented in \citet{PaperI} shows that it is co-located with an asymmetric extension of redshifted CO emission, indicating that it is most likely an infalling stream of material falling on HH~48~NE.
Unfortunately, the tail of HH~48~SW to the South-West is not observed because of the smaller MIRI field of view. 

\begin{figure}[ht]%    \centering
    \includegraphics[width = \linewidth]{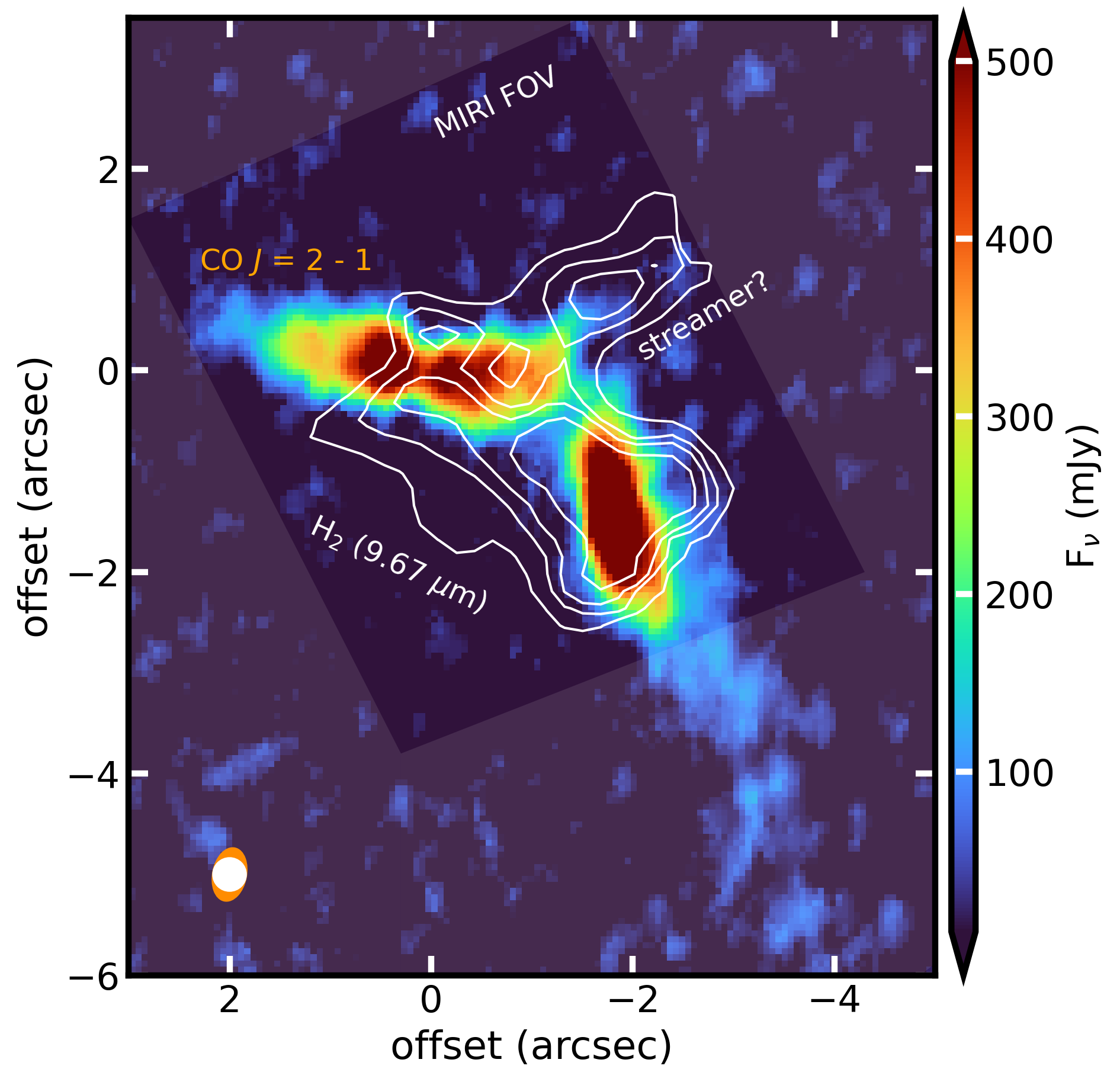}%
    \caption{Overview of the asymmetric emission in the HH~48 system. The colored image shows the integrated CO $J~=~2~-~1$ emission map of HH~48~NE \citep{PaperI}. The contours indicate the spatial extent of the low energy \ce{H2} emission line at 9.67~\micron. The ALMA beam (orange) and MIRI PSF size at 9.7~\micron (white) are shown in the bottom right.}%
    \label{fig:streamer}%
\end{figure}

\section{Spectral extraction}\label{appendix: spectral extraction}
\begin{figure*}[b]%
    \centering
    \subfloat{{\includegraphics[height=5.5cm]{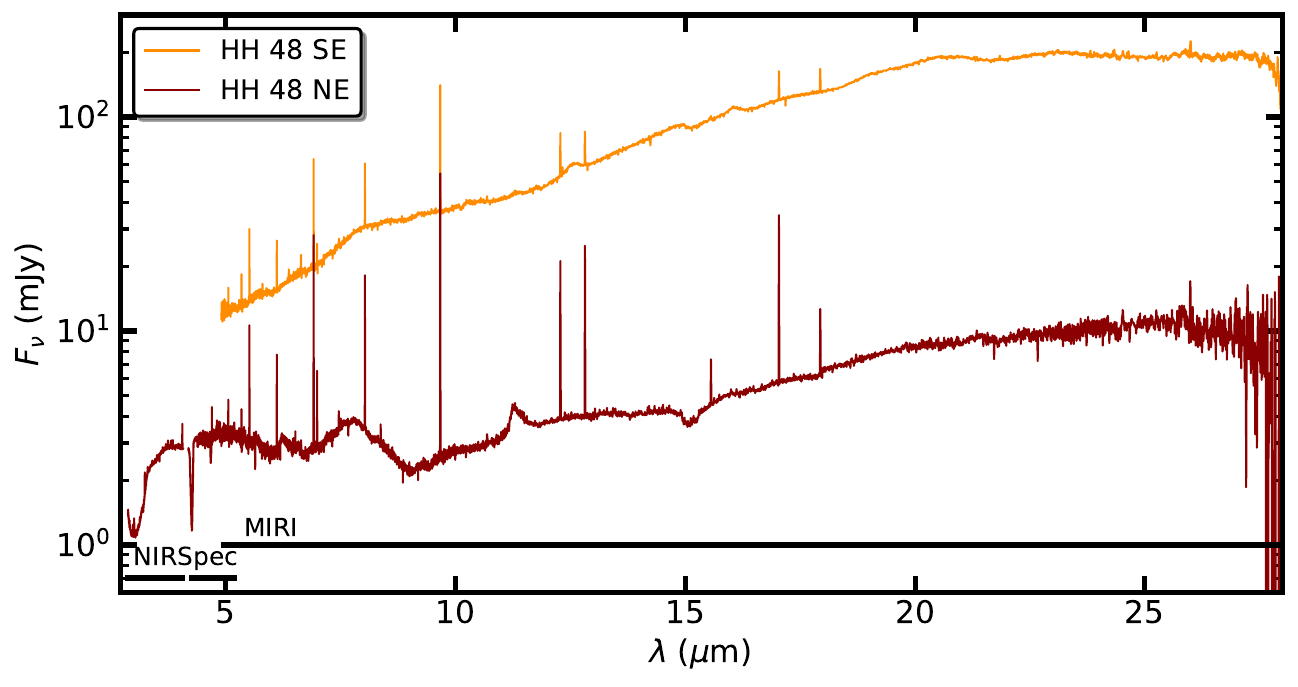}}}%
    \qquad
    \subfloat{{\includegraphics[height=5.5cm]{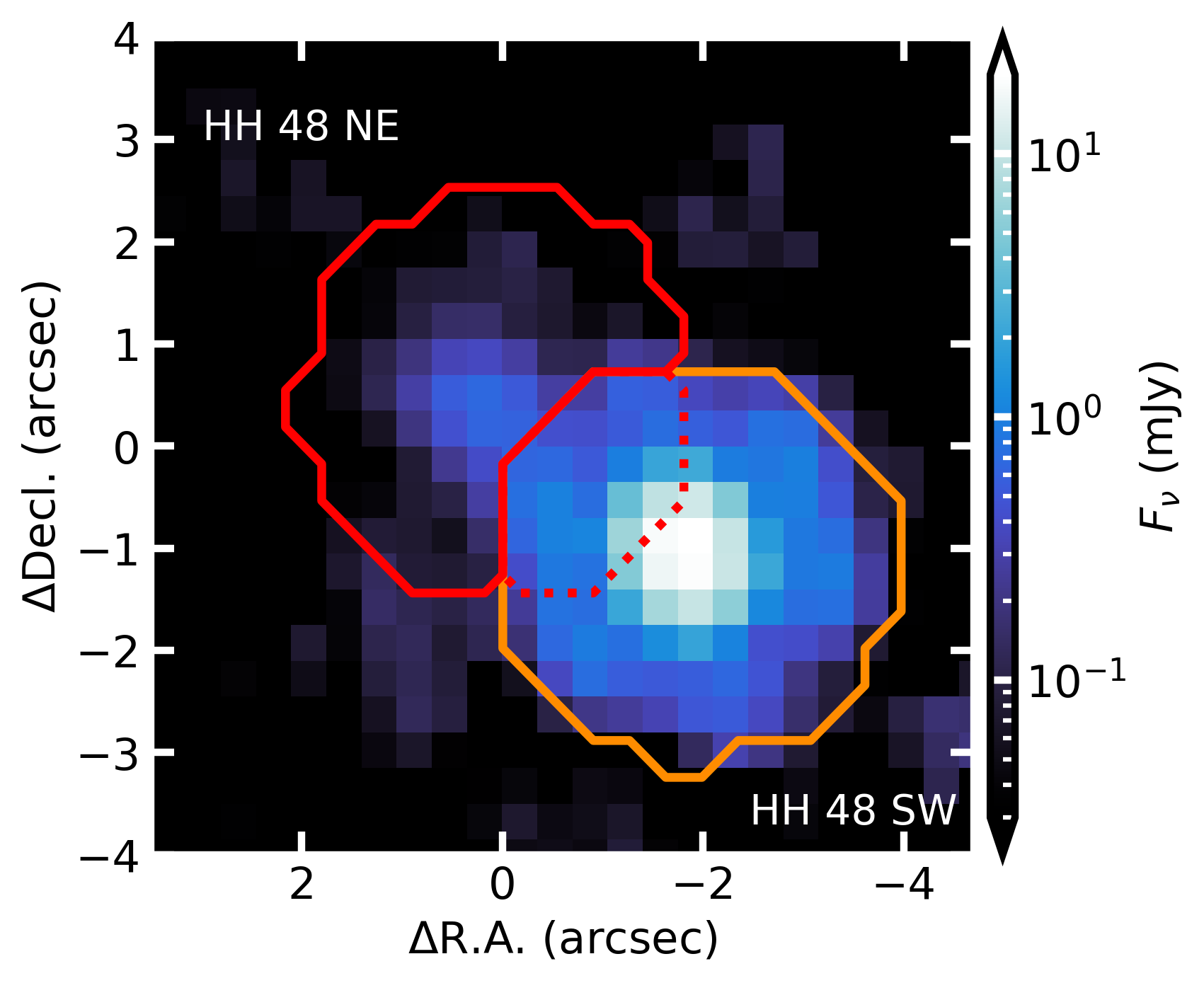}}}%
    \caption{
    Left: The integrated spectrum of HH~48~NE (red) and HH~48~SW (orange). NIRSpec data of HH~48~SW are not available as a result of the smaller field of view of the NIRSpec instrument compared to MIRI.
    Right: Continuum image of the HH~48 system at 24~\micron with the masks used to extract the source integrated spectrum shown on the left.}%
    \label{fig:masks}%
\end{figure*}

The source integrated spectrum is extracted using a conical extraction, increasing the mask with wavelength according to the resolution, to get as much S/N on the spectrum as possible. 
To avoid contamination from the binary component, we used the mask of the primary source, HH~48~SW, as a negative mask on the mask for HH~48~NE (see Fig.~\ref{fig:masks}).

\begin{figure*}[b]%
    \centering
    \includegraphics[width = \linewidth]{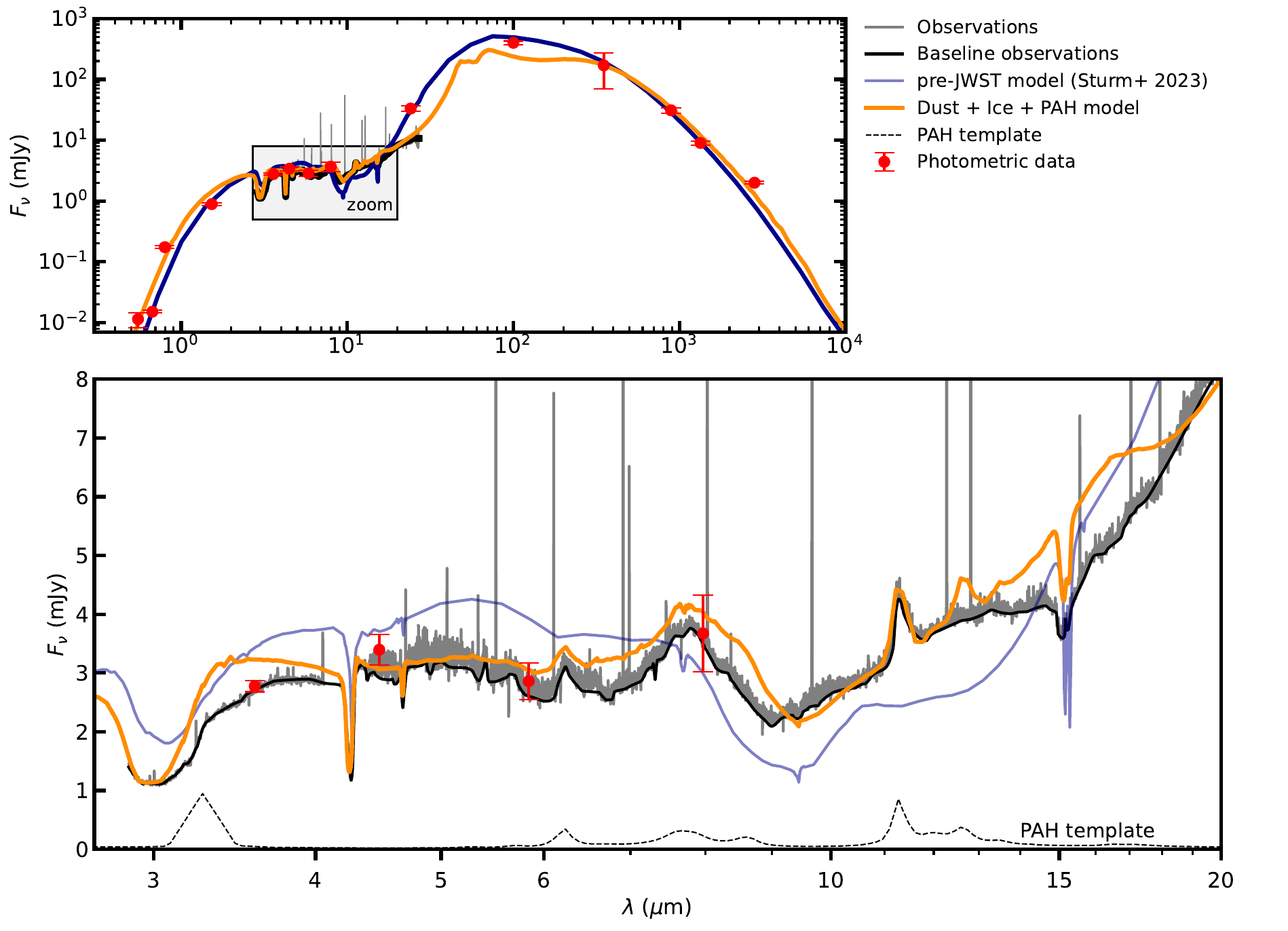}
    \caption{
    Same as Fig.~\ref{fig:SED comparison} but with a comparison between the best fitting model and the pre-JWST model published in \citet{PaperII}}%
    \label{fig:sed comparison sturm2023}%
\end{figure*}

\section{Model improvements}
The current best-fitting model agrees well with the observed spectrum, with most discrepancies being less than 20\%. 
This is an improvement over the model constrained pre-JWST (see Fig.~\ref{fig:sed comparison sturm2023},\ref{fig:2D continuum comparison} for a comparison).
There are a few instances where the model overestimates the flux (by less than 50\%), particularly in the red wing of the water band at 3~\micron. 
This overestimation could be due to an absorption component from ammonia hydrates (\ce{NH3}$\cdot$\ce{H2O}) around 3.5~\micron, which is not accounted for in the model. 
The scattering wing in emission on the red side of the feature resembles the \ce{CO2} feature, further exacerbating the difference. 
The regions around the silicate features (8~--~9~\micron and 13~--~18~\micron) are overestimated by 50\%, possibly due to the source structure and the dust scattering characteristics in the system. 
The disk's edge-on orientation suggests that the silicate features may have a scattered emission component from the warm inner disk and an absorption component from the cold outer disk. 
The ratio of these components is highly sensitive to the source structure, and they may not necessarily reflect similar grain compositions.
The observations extending beyond the model's spatial extent suggest that the model's continuum source is mainly direct dust emission rather than scattered emission, indicating a need for a higher placement of the scattering surface. 
An analysis using a two-component Gaussian fit to the radial profile of the continuum emission reveals a discrepancy in the ratio of scattering to direct emission between the models (1:10) and the observations (1:2). 
While boosting scattering at these wavelengths is not straightforward, it likely involves a combination of source structure and grain composition. 
Investigating the inclusion of a third dust population with micron-sized grains in future studies could offer potential benefits. 
Maintaining a constant grain composition helps constrain the number of free parameters and speed up model computation. 
Future models and JWST observations are expected to provide new insights into grain composition around silicate features, which can be leveraged to improve the model accuracy in the future. 

Improvements can also be made by incorporating foreground cloud absorption in the modeling process. 
While the current model spectrum is scaled with the extinction curve based on $A_\mathrm{V}$, potential ice absorption in an envelope or foreground cloud is not considered. 
In particular models of younger sources may need to account for this. 
Uncertainties exist regarding the distribution of ices over dust grains, which can significantly influence measured abundances. 
The assumed system-averaged gas-to-dust ratio of 100 is consistent with most observations. 
Molecular abundances relative to hydrogen are integrated into the dust grains without altering the total mass in the region, effectively assuming an effective gas-to-solids ratio instead of a gas-to-dust ratio. 
Changes in the mass distribution can alter the physical structure and continuum significantly. 
Therefore, discussing ice-to-rock ratios in relevant disk regions is recommended over abundances when comparing with other systems due to the lack of precise constraints on the vertical gas distribution.

\end{appendix}
\end{document}